\documentclass[]{cit_lab_mfr}
\usepackage{amsmath}
\usepackage{algorithm}
\usepackage{algpseudocode}
\titleformat*{\paragraph}{\normalfont\bfseries}
\newcommand{\sg}{\mathrm{sg}}
\newcommand{\cand}{\mathcal{C}}
\newcommand{\rules}{\mathcal{S}}
\newcommand{\sigmoid}{\operatorname{sigmoid}}
\AtBeginDocument{%
  }
\providecommand{\Description}[1]{}

\title{OneTrans-GSR: A Generative Slate Ranking Paradigm Unifying Reranking and Mix-Ranking with One Autoregressive Transformer}
\author{Global E-Commerce Recommendation Foundation Team}
\affiliation{ByteDance}
\checkdata[Keywords]{Generative Recommendation, Reranking, Mix-Ranking, Slate Generation, Reinforcement Learning, Reward Model}
\date{\today}
\abstract{
Industrial recommender systems allocate substantial computation to pointwise candidate scoring, yet separate reranking and mix-ranking stages still assemble the final exposure slate using hand-crafted rules and specialized low-capacity models. Mix-ranking combines product cards, short videos, livestreams, and advertising cards supplied by distinct upstream business channels. Although these rules and models encode valuable domain knowledge, they are inherently limited in leveraging additional data and computation. They constrain the combinatorial space of candidate slates and determine objective trade-offs through manual tuning. To enable scalable and effective slate generation, we reformulate reranking and mix-ranking as a unified learning problem whose performance can improve with computation. Specifically, we instantiate this formulation across the reranking and mix-ranking stages and propose \textbf{OneTrans-GSR}, a single decoder-only policy that extends our deployed OneTrans backbone to generate the exposure slate end-to-end. During training, a reward model evaluates complete slates and guides reinforcement learning toward slate-level and long-term objectives. The reward model is bounded by offline resources rather than serving-time costs. During decoding, business constraints are explicitly enforced. Offline evaluation shows that slate quality consistently improves as reward-model and policy compute increase. Deployed in a large-scale industrial recommendation system, OneTrans-GSR mixes heterogeneous content types within a unified slate. In online A/B tests, it improves GMV/u by 2.63\%, advertiser value by 2.06\%, and repurchase rate by 2.03\%, while reducing end-to-end latency across the reranking and mix-ranking stages by 13\%. These results mark a paradigm shift in slate generation: moving beyond human-engineered heuristics toward a general learned policy creates a scalable path for further algorithmic optimization and compute-driven improvement.
}

\begin{document}
\maketitle
\section{Introduction}
\label{sec:intro}

Industrial recommender systems are typically organized as cascades that retrieve candidates and refine them with increasingly accurate pointwise scoring models~\cite{covington2016youtube}. The scoring stages of these cascades have benefited from scaling. Architectures such as HSTU, Wukong, and OneTrans consistently translate additional data and computation into improved ranking accuracy~\cite{zhai2024hstu,zhang2024wukong,zhang2025onetrans}. Slate generation, however, has not followed the same trajectory. The slate ultimately exposed to users is still assembled by an expert-engineered system, built around hand-crafted rules and specialized components for tasks such as scattering, quota allocation, de-duplication, fatigue control, and commercial boosting. These are combined with a shallow model that scores candidates under a strict latency budget and with manually tuned fusion weights. Slate generation is a combinatorial problem, and only a small part of the search space is reachable under this design. Extra computation at serving time is therefore spent on a more exhaustive search over this restricted space, and expanding the space requires additional rules and components. Slate quality thus improves with expert knowledge and engineering effort rather than with data and computation.

\begin{figure}[t]
  \centering
  \includegraphics[width=0.85\linewidth]{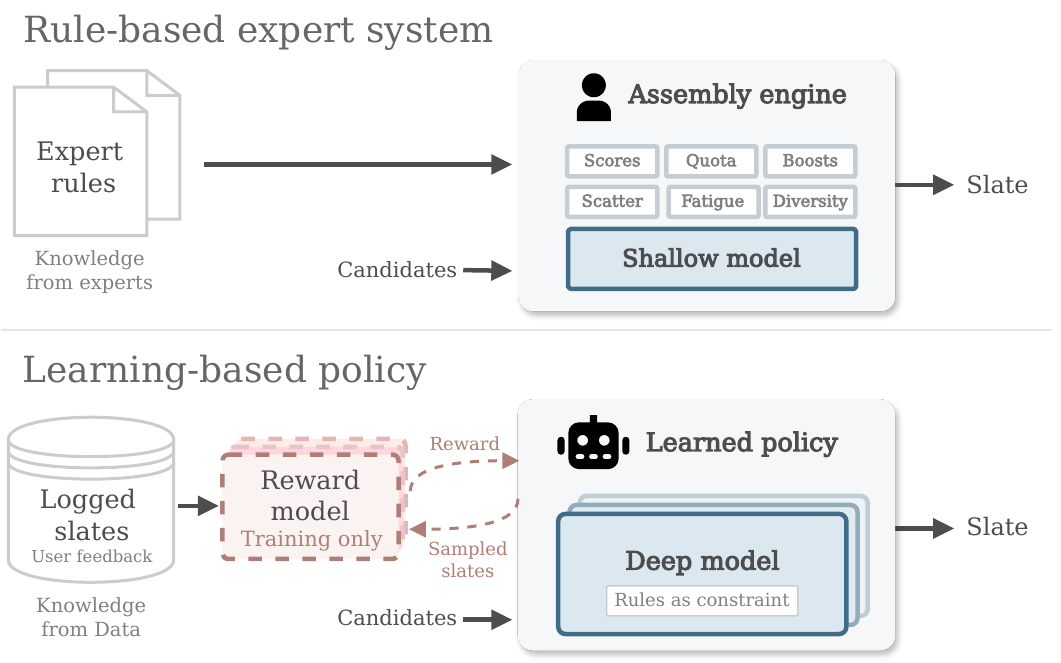}
  \caption{Two paradigms for slate generation. The rule-based expert system executes expert rules around a shallow model; the learning-based policy learns from logged slates through a reward model used only in training, and a single deep model generates the slate, with rules retained only as auxiliary constraints.}
  \Description{Two-row diagram. The upper row shows expert rules feeding an assembly engine that contains rule modules and a shallow model, which outputs a slate. The lower row shows logged user feedback training a reward model, which in turn trains a deep learned policy that outputs a slate with rules applied as constraints.}
  \label{fig:paradigm}
\end{figure}

We instead formulate slate generation as a learning problem. A reward model is trained on logged user feedback to evaluate complete slates, so the objectives that shape a slate are learned from data rather than encoded by hand. A single general policy is then optimized against this reward to generate slates from the candidate pool, and the search space is no longer restricted by hand-crafted rules but covers all feasible slates over the candidates. The reward model is used only during training, and its capacity is therefore limited by offline resources rather than by the serving budget. With a larger reward model and a deeper policy, additional data and computation now translate into better slates, exactly the scaling behavior that the expert-engineered system lacks. What changes is the role of the rules rather than their presence. Hard constraints are enforced during decoding, and soft preferences are incorporated into the reward. Figure~\ref{fig:paradigm} contrasts the rule-based expert system with the learning-based policy. 

We instantiate the unified reranking and mix-ranking formulation in a large-scale industrial recommendation system for e-commerce. Candidates---product cards, short videos, livestreams, and advertising cards---are supplied by distinct upstream business channels and compete for positions within a single exposure slate under multiple objectives and business constraints. Recent advances provide the necessary mechanisms. Reinforcement-learning post-training demonstrates that autoregressive policies can be optimized directly against sequence-level rewards~\cite{ouyang2022instructgpt,shao2024grpo}. Generative recommendation systems similarly cast retrieval and ranking as sequential generation~\cite{rajput2023tiger,zhai2024hstu,deng2025onerec}, and reported studies show performance gains with model scale and training compute~\cite{zhai2024hstu,deng2025onerec}. Instantiating the formulation in production, however, introduces three challenges. First, unlike language generation over a fixed vocabulary, slate generation selects from a request-specific pool of pre-scored candidates. Second, reward is defined over the complete slate rather than individual items. An item's utility depends jointly on its position, neighboring items, and the slate's composition across content types. Slate-level feedback is therefore non-decomposable into independent item-level supervision. Third, every generated slate must satisfy production business constraints, whereas unconstrained autoregressive generation provides no such guarantees. These challenges lead to the central question of this work: can a single learned policy, optimized offline against slate-level rewards, replace the expert-engineered system that assembles the slate today?

\textbf{OneTrans-GSR} (\textbf{G}enerative \textbf{S}late \textbf{R}anking) answers this question in production. The system models slate generation as autoregressive decoding over the per-request candidates. A single decoder receives them from the upstream fine-ranker and emits the exposure slate directly, so no separate rule-based assembly engine, specialized components, or downstream reranking remain on the serving path. The decoder is the only model served at this stage. The training signal, in turn, comes from a slate-level reward model that scores complete slates rather than summing independent item scores. The decoder and the reward model are both built from OneTrans blocks~\cite{zhang2025onetrans} and use the fine-ranking score as a candidate feature. OneTrans thus serves as a shared backbone across fine ranking, reward modeling, and slate generation. Deployed at full traffic, OneTrans-GSR paves the way for gradually integrating fine ranking into slate generation.

We train the decoder in two stages. The decoder is first pretrained on long user-behavior sequences with a next-token prediction loss, and then post-trained with reinforcement learning (RL) against the reward model. For the post-training stage we propose Geometric-Mean Reward Policy Optimization (GMR-PO), a method for multi-objective policy optimization. GMR-PO aggregates the reward model's scores for multiple objectives into a scalar reward. The scalar reward and the reward for business preferences are standardized independently, and the resulting advantages are combined to optimize the policy. Soft business preferences thus serve as training signals, while advertiser value and hard business constraints are controlled during decoding.

In summary, we formulate slate generation as a learning problem and validate the resulting policy in production. Our contributions are as follows:
\begin{itemize}
  \item \textbf{A unified generative policy for reranking and mix-ranking.} We formulate slate generation over heterogeneous, request-specific candidate pools as a constrained autoregressive policy. OneTrans-GSR jointly handles candidate selection, content-type mixing, and business-aware ordering, while decoupling slate-level evaluation from online serving to eliminate the need for a separate online evaluator or downstream reranking module.
  \item \textbf{Slate-level reward modeling.} We introduce a reward model that scores complete slates rather than aggregating per-item predictions, providing slate-level signals for conversion and long-term objectives. Used only at training time, its capacity is bounded by offline resources rather than serving costs, so it can be much larger than any model that could be served online.
  \item \textbf{Multi-objective and rule-aware post-training.} We develop GMR-PO to jointly optimize utility and business objectives. GMR-PO aggregates the reward model's scores across objectives into a scalar reward. The scalar reward and a separate reward encoding soft business preferences are standardized separately in the advantage computation. The policy learns how these objectives affect slate construction within a
  unified optimization procedure, and business requirements are satisfied within generation rather than repaired by downstream reordering.
  \item \textbf{Production validation and compute scaling.} We validate OneTrans-GSR through controlled offline studies, scaling analyses, system benchmarks, and production A/B tests. Against a strong Generator--Evaluator baseline~\cite{feng2021prs,shi2023pier}, the deployed system lifts GMV/u by 2.63\%, advertiser value by 2.06\%, and repurchase rate by 2.03\%, while reducing end-to-end latency across the reranking and mix-ranking stages by $13\%$. In offline evaluation, slate quality improves consistently with reward-model, decoder, and rollout compute.
\end{itemize}
\section{Related Work}
\label{sec:related}

\noindent\textbf{Reranking, Mix-Ranking, and Slate Generation} Reranking takes the candidates scored by the upstream ranker and decides which items are shown and in what order. Mix-ranking combines product cards, short videos, livestreams, and advertising cards from distinct upstream business channels into a single exposure slate under business constraints. Existing approaches differ mainly in the extent to which these decisions are learned. Early methods retain the candidate set and update item scores. MMR selects items by balancing relevance against redundancy~\cite{carbonell1998mmr}, while context-aware rankers condition each candidate's score on the other items in the list, as in DLCM~\cite{ai2018dlcm}, RankFormer~\cite{buyl2023rankformer}, MIR~\cite{xi2022mir}, and PRM~\cite{pei2019prm}. A further step is to optimize the slate as a whole: Exact-k treats slate selection as a combinatorial problem~\cite{gong2019exactk}, Seq2Slate learns an autoregressive ranking policy from click logs~\cite{bello2018seq2slate}, and SlateQ decomposes long-term slate value to make reinforcement learning tractable~\cite{ie2019slateq}. Beyond one-step CTR, Liu and Liu optimize expected session clicks through reinforcement learning while accounting for user departure~\cite{liu2023ctr}. Closer to our setting, the generator itself is learned and served. GRN trains a generator by policy gradient against an evaluator used only during training and deploys the generator at Taobao~\cite{feng2021grn}, while NAR4Rec generates the slate non-autoregressively~\cite{ren2024nar4rec}; Section~\ref{sec:decoder-design} compares autoregressive and non-autoregressive decoding in our setting. Some production systems instead retain an online evaluator: PRS~\cite{feng2021prs} and PIER~\cite{shi2023pier} select the final slate from alternatives proposed by separate generators. OneTrans-GSR replaces the separate reranking, mix-ranking, and online evaluation stages with a single policy that generates the final slate directly, while slate evaluation is confined to training.

\noindent\textbf{Generative Recommendation.}
Generative recommendation formulates retrieval or ranking as sequence generation. Early work focuses on item generation: retrieval based on semantic IDs decodes item identifiers~\cite{rajput2023tiger}, while session generation extends this approach to entire sessions~\cite{deng2025onerec}. Scalable sequence backbones for generative recommendation include HSTU and OneTrans~\cite{zhai2024hstu,zhang2025onetrans}. Generative rerankers instead produce complete slates, with designs varying along two axes. The first is the decoding scheme: autoregressive, non-autoregressive, or hierarchical. The second is the training signal: preference alignment, supervision from an offline evaluator, or rewards defined over multiple slate-level objectives~\cite{lin2025gref,wang2025nlgr,song2026degre,wang2026nsgr,meng2025sortgen,pang2026higr}. Recent work advances both axes. On the decoding side, methods seek to retain the quality of autoregressive generation at lower inference cost through distillation into non-autoregressive models, confidence-guided denoising, and graph-based slate generation~\cite{zhang2026dualrerank,jia2026unirank,yang2026congrats}. On the training side, recent methods explore group-relative objectives, globally consistent action spaces based on semantic IDs, and objectives that account for constraints or exploration~\cite{zhang2026goalrank,jia2026glorank,li2026constraintaware,zhao2026degr}. OneTrans-GSR combines autoregressive decoding with slate-level reward learning to unify reranking and mix-ranking over a heterogeneous candidate pool. A single decoder, trained against an offline reward model, serves the slate directly without an online evaluator or downstream reranker. Business requirements enter the same generation process: soft preferences provide training rewards, while strict constraints are applied during decoding.

\noindent\textbf{Preference Alignment and Multi-Objective Policy Optimization.} Optimizing a policy against learned rewards raises two questions: how to update the policy and how to combine multiple rewards. Policy updates have been studied extensively in language modeling. In the standard setting for preference alignment, a reward model is trained from preference data, and the policy is subsequently optimized against its scores~\cite{christiano2017rlhf,ouyang2022instructgpt}. Direct preference methods avoid an explicit reward model and optimize policies directly from preference data, as in DPO~\cite{rafailov2023dpo} and its extensions IPO, SimPO, and S-DPO~\cite{azar2024ipo,meng2024simpo,chen2024sdpo}. Group-relative methods instead estimate advantages by comparing outputs sampled for the same input. GRPO computes advantages within a group of sampled outputs~\cite{shao2024grpo}, while GSPO and GMPO differ in how they aggregate importance ratios or per-token terms~\cite{zheng2025gspo,zhao2025gmpo}. Existing alignment methods therefore differ primarily in the policy update rather than in the treatment of multiple reward dimensions.

The second question concerns multiple objectives. Reinforcement learning commonly reduces multiple objectives to a scalar reward, an approach also used in e-commerce recommendation~\cite{hayes2022morl,lin2019pareto} and, more recently, in preference optimization. GDPO standardizes each reward dimension before aggregation, reducing dominance caused by differences in raw reward scales~\cite{gdpo2026}. SAGE separates the multi-objective advantage from the importance ratio and computes the latter geometrically over the sequence~\cite{xie2026sage}. OneTrans-GSR uses rewards defined over complete slates and combines short- and long-term utility objectives with soft business preferences. GMR-PO aggregates the reward model's scores across utility objectives into a scalar reward. The scalar reward and a separate reward encoding soft business preferences are standardized independently before the resulting advantages are combined. Policy optimization uses a clipped surrogate objective with KL regularization on fresh rollout batches, without additional correction for off-policy samples (Section~\ref{sec:posttrain}).

\section{Problem Formulation}
\label{sec:problem}

For each request, OneTrans-GSR receives a pool $\cand=\{c_1,\dots,c_{N_c}\}$ of $N_c$ candidates. The heterogeneous candidate pool contains product cards, short videos, livestreams, and
advertising cards supplied by distinct upstream business channels. Conditioned on the user and context features $u$, the policy selects $T$ candidates from $\cand$ without replacement. The
selected candidates form an ordered exposure slate $L=(c_{i_1},\dots,c_{i_T})$. The policy factorizes the probability of the slate autoregressively:
\begin{equation}
\pi_\theta(L \mid u, \cand)
= \prod_{t=1}^{T}
\pi_\theta\big(c_{i_t} \mid c_{i_{<t}}, u, \cand\big),
\label{eq:policy}
\end{equation}
where $c_{i_{<t}}=(c_{i_1},\dots,c_{i_{t-1}})$ denotes the candidates selected before step $t$. We optimize the policy parameters $\theta$ to maximize the expected utility of the complete
slate:
\begin{equation}
\max_{\theta}\;
\mathbb{E}_{L \sim \pi_\theta(\cdot\mid u,\cand)}
\left[\mathcal{R}(L,u)\right],
\label{eq:objective}
\end{equation}
where $\mathcal{R}$ denotes the underlying utility of slate $L$ for context $u$. In practice, the reward model's scores for different objectives are normalized and aggregated into a scalar
reward. The scalar reward and an additional reward encoding soft business preferences are standardized independently before the resulting advantages are combined according to
Eq.~\eqref{eq:rulesep}. Strict constraints are not included in $\mathcal{R}$; they instead restrict the decoding distribution. The reward and decoding mechanisms jointly address category
diversity, exposure quotas, deduplication, fatigue control, and cold-start exposure (Section~\ref{sec:constraints}).

\section{System Overview}
\label{sec:system-overview}

\begin{figure}[!t]
\centering
\includegraphics[width=\linewidth]{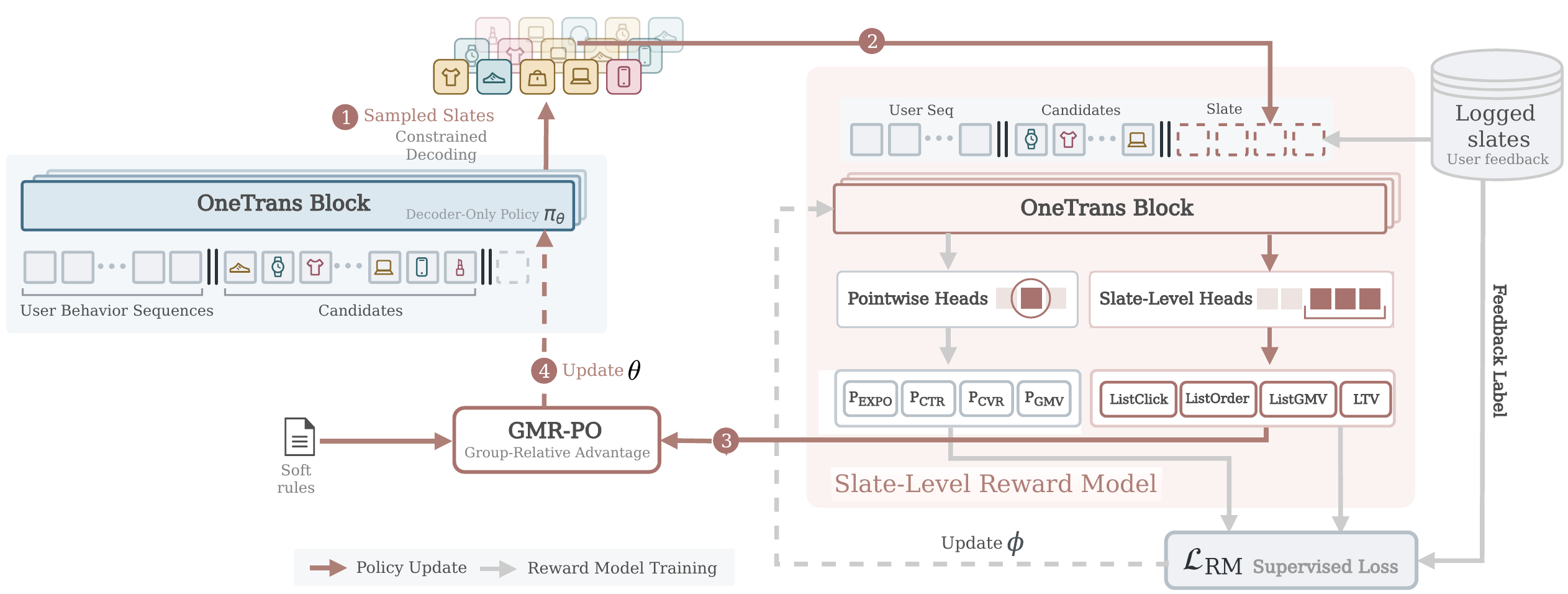}
\caption{Offline training pipeline of OneTrans-GSR. ListClick, ListOrder, and ListGMV denote the total click count, total order count, and total gross merchandise value within a slate,
respectively. The reward model and decoder are pretrained separately. During policy post-training, (1) the decoder generates constrained slate samples, (2) a frozen RM ensemble evaluates the
samples, (3) GMR-PO converts the model scores and a reward for soft business preferences into advantages, and (4) the resulting advantages update the decoder.}
\label{fig:training}
\end{figure}

\begin{figure}[!t]
\centering
\includegraphics[width=0.6\linewidth]{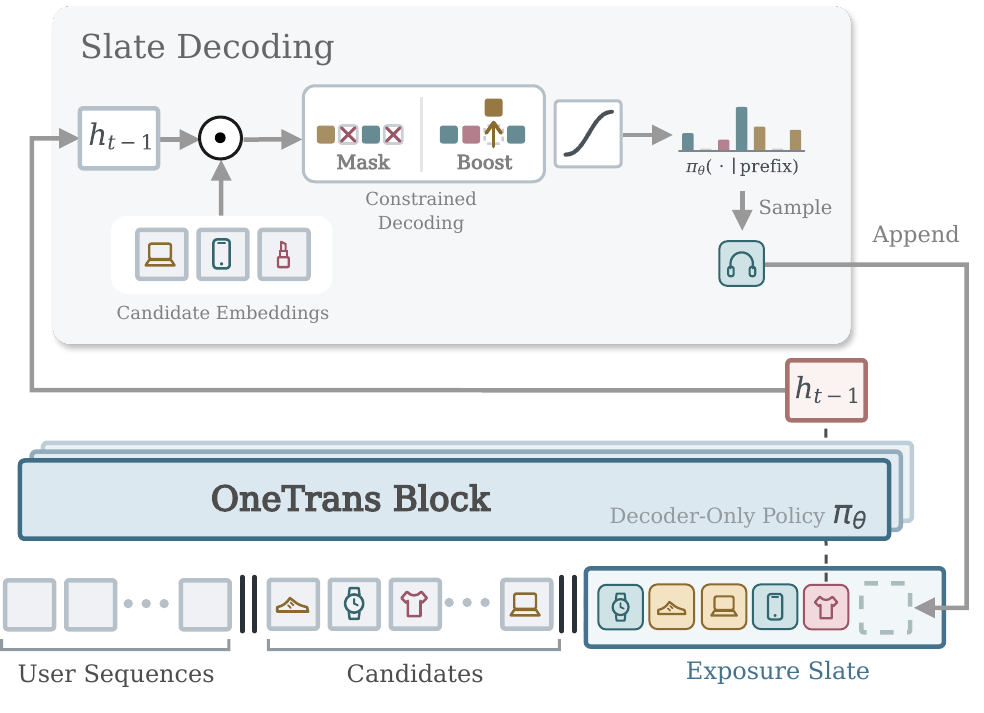}
\caption{Decoder-only online serving pipeline of OneTrans-GSR. A prefill pass initializes the KV cache from the user behavior and candidate tokens. During autoregressive decoding, masks and additive logit adjustments apply hard constraints and soft controls. Multiple slates are decoded in parallel, and the slate with the highest accumulated sequence log-probability is served directly.}

\label{fig:serving}
\end{figure}

Figures~\ref{fig:training} and~\ref{fig:serving} illustrate the offline training and online serving pipelines of OneTrans-GSR, respectively. The system comprises a slate-level reward model (RM; Section~\ref{sec:rm}) and an autoregressive decoder (Section~\ref{sec:decoder}), both built from OneTrans blocks~\cite{zhang2025onetrans} with a shared candidate-token representation but independent parameters. The RM provides slate-level supervision during training, while only the decoder is deployed online.

\noindent\textbf{Training.} The training pipeline consists of three stages.

\noindent\textit{Stage 1: Reward Model Pretraining.} The RM learns to evaluate complete slates from logged exposures and user feedback, with auxiliary supervision from pointwise and content-type preference heads (Section~\ref{sec:rm}).

\noindent\textit{Stage 2: Decoder Pretraining.} The decoder is pretrained on long user behavior sequences using next-token prediction and subsequently adapted to generate slates from request-specific candidate pools.

\noindent\textit{Stage 3: Policy Post-Training.} The decoder generates constrained slate samples for evaluation by a frozen RM ensemble. GMR-PO aggregates multi-objective RM scores into a scalar reward and combines its standardized advantage with that of a separate reward for soft business preferences. The combined advantage drives policy optimization with KL regularization (Section~\ref{sec:posttrain}).

\noindent\textbf{Serving.} Each request supplies user behavior sequences, a heterogeneous candidate pool, and business-rule inputs. A single prefill pass contextualizes the user and candidate tokens and initializes a KV cache for subsequent decoding. At each autoregressive step, the decoder scores the candidates, applies masks and additive logit adjustments to enforce hard constraints and guide soft business controls, and samples the next item from the resulting distribution. Generation continues until the requested slate length is reached.

OneTrans-GSR decodes \(G_{\mathrm{serve}}\) candidate slates in parallel and serves the one with the highest accumulated sequence log-probability, without invoking the RM or a separate downstream reranking or mix-ranking module.

\section{Method}
\label{sec:method}

OneTrans-GSR learns to generate business-constrained slates through reward-guided policy optimization. A slate-level reward model provides multi-objective training signals, while GMR-PO integrates these signals with soft business preferences to optimize the autoregressive decoder.

\subsection{Slate-Level Reward Model}
\label{sec:rm}

The RM has two components. The \emph{efficiency RM}, parameterized by $\phi$, outputs $\mathbf{z}_\phi(L,u)\in\mathbb{R}^{3}$ for slate $L$ and context $u$. The three logits correspond to
ListClick (the total click count within a slate), ListOrder (the total order count within a slate), and ListGMV (the total gross merchandise value within a slate). The \emph{LTV RM} learns
from delayed repurchase labels (Section~\ref{sec:rm-ltv}). Each candidate is represented by one token that combines user, item, context, and dense features. The decoder uses the same
representation for candidate tokens but has independent parameters. The upstream fine-ranking score is included as an input feature, and Section~\ref{sec:listwise-interaction} reports its
ablation.

\subsubsection{Slate-Level Modeling with an EOS Token}
\label{sec:rm-listwise}

We append a learnable EOS token after the item tokens to score the complete slate. The final hidden state $h_{\mathrm{EOS}}$ provides the logits for the three objectives:
\begin{equation}
\mathbf{z}_\phi(L,u) = \big(z^{\mathrm{click}},\, z^{\mathrm{order}},\, z^{\mathrm{gmv}}\big) = f_\phi\big(h_{\mathrm{EOS}}\big).
\label{eq:rm-readout}
\end{equation}

Under causal attention, each item token attends to itself and the preceding items, while the terminal EOS token attends to the complete slate. The EOS state $h_{\mathrm{EOS}}$ therefore
summarizes the slate in exposure order without allowing an item representation to depend on subsequent items.

User feedback on logged exposure slates provides supervision. For each objective $v\in\mathcal{V}=\{\mathrm{click},\mathrm{order},\mathrm{gmv}\}$, let $Y_v(L)$ denote the total feedback for slate $L$, such as its click count, and let $o_v(L)=\mathbb{I}[Y_v(L)>0]$ indicate whether the slate is positive. We model all three objectives with weighted logistic regression (WLR)~\cite{covington2016youtube}. WLR classifies $o_v$ and assigns larger weights to positive slates with higher values:
\begin{equation}
\begin{gathered}
\mathcal{L}_{\mathrm{list}}
= \sum_{v\in\mathcal{V}} \omega_v(L)\;
  \mathrm{BCE}\big(o_v(L),\, \sigmoid(z^v(L,u))\big), \\
\omega_v(L)=
\begin{cases}
  a_v(Y_v(L)) & \text{if } o_v(L)=1,\\
  1 & \text{otherwise},
\end{cases}
\end{gathered}
\label{eq:rm-wlr}
\end{equation}

Here, $\mathrm{BCE}$ denotes binary cross-entropy, and $a_v(\cdot)>0$ determines the weight assigned to a positive example. The weights for click and order are the corresponding slate counts, while the GMV weight is the clipped and normalized slate GMV. An auxiliary head at each item token uses the same WLR loss with item-level labels. The resulting per-token losses form the
auxiliary loss $\mathcal{L}_{\mathrm{point}}$.

Because slate labels are sums of item labels, an alternative readout estimates slate utility by summing item-level predictions. The EOS readout instead receives direct supervision from the slate label and learns from the ordered slate. For policy post-training, we transform each logit into a positive reward score via $R^v(L,u)=\exp(z^v(L,u))$. The transformation provides a monotonic score for relative comparisons among rollouts after normalization; $R^v$ is not interpreted as a calibrated expectation of click count, order count, or GMV. The per-item heads provide auxiliary supervision and monitoring. Section~\ref{sec:listwise-interaction} compares the two readouts using calibration and controlled slate perturbations.

\subsubsection{Extending the EOS Readout to Preferences over Content Types}
\label{sec:rm-heads}

A dedicated token models user preferences over the content types $\mathcal{K}=\{\mathrm{product},\mathrm{live},\mathrm{video}\}$. Advertising candidates are excluded from this auxiliary task and controlled separately during decoding. Mix-ranking uses the learned preferences to allocate exposure among the content types in $\mathcal{K}$. We add to each item token a feature $ \mathbf{f}^{\mathrm{type}}_u$ that summarizes the user's exposure, click, and conversion distributions across these content types. A preference token $e_{\mathrm{pref}}$ is appended to the sequence, and the token's final state $h_{\mathrm{pref}}$ is passed to one prediction head for each pair of content type and objective.

Head $\psi_{k,v}$ predicts objective $v$ for items of content type $k$:
\nopagebreak[4]
\begin{equation}
\hat p_{k,v}(L,u)=
\sigmoid\!\left(\psi_{k,v}(h_{\mathrm{pref}})\right),
\qquad (k,v)\in\mathcal{K}\times\mathcal{V}.
\label{eq:content-type}
\end{equation}

For example, $\hat p_{\mathrm{video},\mathrm{click}}$ is a weighted prediction score for clicks on videos in the slate. Each head uses the WLR loss in Eq.~\eqref{eq:rm-wlr}, with labels computed only from items of the corresponding content type. Let $n_k(L)$ denote the number of impressions of content type $k$ in slate $L$. We weight the corresponding loss by $1/
\sqrt{n_k(L)}$ to reduce imbalance across content types. Losses for content types absent from the slate are masked. The prediction heads provide auxiliary supervision for the shared RM backbone.

The efficiency RM objective combines the main and auxiliary losses:
\begin{equation}
\mathcal{L}_{\mathrm{RM}} = \mathcal{L}_{\mathrm{list}} + \mathcal{L}_{\mathrm{point}} + \mathcal{L}_{\mathrm{type}}.
\label{eq:rm-loss}
\end{equation}
Section~\ref{sec:listwise-interaction} evaluates the EOS readout and the component that models preferences over the content types in $\mathcal{K}$.

\subsubsection{LTV Reward Model}
\label{sec:rm-ltv}

The LTV RM predicts repurchase within 14 days as an offline proxy for long-term value. Because the label is delayed, the model uses the same OneTrans architecture as the efficiency RM but has separate parameters and embeddings. Delayed updates therefore do not affect the efficiency RM. Each training example contains only features available at request time and is used only after its repurchase label is fully observed.

Because repurchase events are too sparse and delayed for the slate-level training described in Section~\ref{sec:rm-listwise}, we aggregate pointwise predictions into a slate-level LTV reward. The aggregation follows the item-wise value decomposition of SlateQ~\cite{ie2019slateq} and assumes that a user consumes at most one item per slate. Production logs support this approximation: 98.3\% of slates contain at most one click, and 99.99\% contain at most one purchase.

Three predictions represent the sequence from click to repurchase. $\mathrm{pCTR}(c,u)$ is the probability that item $c$ is clicked, $\mathrm{pCVR}(c,u)$ is the probability that the click leads to a purchase, and $\mathrm{pREP}(c,u)$ is the probability that the purchase is followed by a repurchase within 14 days. The efficiency RM supplies $\mathrm{pCTR}$ and $\mathrm{pCVR}$, while the LTV RM supplies $\mathrm{pREP}$. The slate-level LTV reward sums this sequence over the items:
\begin{equation}
R^{\mathrm{ltv}}(L,u) = \sum_{c \in L} \mathrm{pCTR}(c,u)\cdot \mathrm{pCVR}(c,u)\cdot \mathrm{pREP}(c,u).
\label{eq:ltv}
\end{equation}

The efficiency RM learns from immediate feedback, while the LTV RM learns from a separate delayed stream containing only completed repurchase labels. Incomplete labels do not enter the LTV loss. GMR-PO geometrically aggregates the LTV score with the click, order, and GMV scores (Section~\ref{sec:posttrain}).

\subsection{Autoregressive Decoder}
\label{sec:decoder}

The decoder uses OneTrans blocks to parameterize the policy in Eq.~\eqref{eq:policy}. The input comprises two segments: the \emph{prefilled context}, which contains the user behavior
sequences and candidate pool, and the \emph{generated sequence}, which contains the items selected autoregressively.

\subsubsection{Prefill}
\label{sec:decoder-prefill}

We place the user behavior sequences and all $N_c$ candidate tokens before the generated slate (Figure~\ref{fig:serving}). During prefill, candidates interact with the user context and one another through a block-causal attention mask. Let $H$, $C$, and $P_{\mathrm{gen}}$ denote the sets of token positions for user behavior, candidates, and generated slate tokens, respectively. Tokens at positions in $H$ retain the masks of their original behavior sequences; tokens at positions in $C$ attend to all tokens in $H \cup C$; and tokens at positions in $P_{\mathrm{gen}}$ attend to $H \cup C$ and earlier generated positions. Prefilled tokens never attend to generated positions. Each candidate representation is therefore conditioned on the user and the complete candidate pool before generation begins, while slate generation remains autoregressive.

Prefill allows candidates to interact from the first layer. By contrast, output-only scoring compares candidates only at the final dot-product head, while cross-attention introduces projected candidate representations within the decoder layers. Section~\ref{sec:decoder-design} compares the three designs. Although prefill lengthens the input sequence, its computation is performed once per request and shared across decoding steps.

\subsubsection{Slate Generation}
\label{sec:decoder-gen}

The decoder selects items from the candidate pool without replacement. A SEP token is appended after the prefilled context and used as the BOS token. The hidden state $h_0$ of the SEP token is used to select the first item. Let $c_{<t}=(c_{i_1},\dots,c_{i_{t-1}})$ denote the prefix generated before step $t$. At step $t$, the prefilled context and generated prefix produce the hidden state $h_{t-1}$. Let $\mathbf e_i$ denote the embedding of candidate $c_i$ in the table $\mathbf{E}_\cand$. We compute a temperature-scaled cosine score, apply logit adjustments, and take a softmax over the candidates:

\begin{equation}
\begin{gathered}
s_{t,i} = \frac{\cos\bigl(h_{t-1},\, \sg(\mathbf e_i)\bigr)}{\tau},
\quad
z_{t,i} = s_{t,i} + \delta_t(c_i) + m_{t,i}, \\
\pi_\theta(c_i \mid c_{<t},u,\cand)
= \frac{\exp(z_{t,i})}{\sum_{j=1}^{N_c}\exp(z_{t,j})},
\end{gathered}
\label{eq:decoder-policy}
\end{equation}

The operator $\sg(\cdot)$ denotes stop-gradient, as explained below. The adjustment $\delta_t(c_i)$ applies business controls (Section~\ref{sec:constraints}), while the availability mask $m_{t,i}$ is $0$ for available candidates and $-\infty$ for candidates already selected. Following CLIP~\cite{radford2021clip}, we learn the inverse temperature as $1/ \tau=\min(\exp(\beta_\tau),100)$. The upper bound limits the logit scale. Training rollouts and serving use the same distribution at each decoding step.

Unlike a language model with a fixed vocabulary, OneTrans-GSR operates over a large and continuously changing item corpus. Each request supplies its own candidate pool, whose embeddings form an output vocabulary specific to that request. Equation~\eqref{eq:decoder-policy} scores this vocabulary against the decoder state.

Post-training, offline evaluation, and serving use the same sampling procedure but may use different sampling budgets. During post-training, the decoder samples $G_{\mathrm{train}}$ slates per request, and the sampled slates form the rollout group for GMR-PO (Section~\ref{sec:posttrain}). Offline evaluation and serving sample $G_{\mathrm{serve}}$ slates and select the slate with the highest sequence log-probability. The selected slate is served directly without RM-based selection or separate downstream reranking or mix-ranking.

\noindent\textbf{Gradient Isolation for the Candidate Embeddings.}
\label{sec:decoder-sg}
The candidate table $\mathbf{E}_\cand$ provides both the input embeddings for prefill and the output vocabulary for candidate selection. Using the same table for input encoding and output scoring creates two distinct gradient paths. Gradients from the input path encourage the embeddings to encode item semantics. For positive advantages, gradients from the output path can pull selected candidates toward the decoder state and push unselected candidates away; negative advantages reverse these directions.

Gradients from the output path may become correlated across requests. Frequently selected items from the same category can consequently receive similar updates, reducing embedding diversity and contributing to collapse. Embedding collapse has also been reported when scaling recommendation models~\cite{guo2024collapse}.

We stop the gradient through the candidate vector on the output path, including its normalization, in Eq.~\eqref{eq:decoder-policy}. The table $\mathbf{E}_\cand$ is therefore updated only through the input path. Section~\ref{sec:decoder-design} reports the corresponding ablation. The separation resembles the distinct input and output embeddings used by word2vec~\cite{mikolov2013word2vec}, while SimSiam similarly uses stop-gradient to prevent collapse~\cite{chen2021simsiam}.

\subsubsection{KV Cache for Parallel Rollouts}
\label{sec:decoder-kv}

Parallel generation maintains two KV caches. The \emph{prefill cache} stores states for the user behavior sequences and candidate pool, is computed once per request, and is shared across parallel samples. Each sample maintains a private \emph{autoregressive cache} for its generated tokens. The autoregressive cache grows by one token at each step, and parallel samples remain independent without pruning or reordering.

The prefilled context accounts for most of the sequence length, so replicating it for every sample would dominate cache memory. Sharing the prefill cache keeps peak GPU memory about 17\% above the baseline without candidate prefill. The two caches support both training rollouts and online serving. Relative to the preceding BF16 configuration, KV caching improves training throughput by \(2.09\times\) and serving throughput by \(4.52\times\). The corresponding cumulative speedups over the eager baseline are \(4.85\times\) and \(11.57\times\), respectively (Table~\ref{tab:systems-efficiency}).

\subsection{Incorporating Business Logic}
\label{sec:constraints}

In industrial systems, reranking and mix-ranking must balance utility objectives with business requirements. During decoding, request-specific masks and finite logit adjustments control spacing across content types and categories, product deduplication, exposure quotas, and advertising exposure (Section~\ref{sec:decode-intervention}). During post-training, a separate reward encourages fatigue reduction, cold-start exposure, content discovery, and diversity (Section~\ref{sec:reward-combine}).

\subsubsection{Constrained Decoding}
\label{sec:decode-intervention}

We incorporate business rules through constrained decoding, following guided generation~\cite{hokamp2017gbs,post2018dba,yang2021fudge}. Let $\rules=\{r_1,r_2,\dots\}$ denote the rules for a request. At step $t$, the adjustment $\delta_t$ in Eq.~\eqref{eq:decoder-policy} sums their contributions:
\begin{equation}
\delta_t(c)=\sum_{r\in\rules}f_{r,t}(c;c_{<t},u),
\label{eq:decode-intervene}
\end{equation}
where $f_{r,t}$ adjusts the logit of candidate $c$ using the generated prefix and user context. Training rollouts and serving apply the same functions. Penalty levels can differ across business controls. We implement two forms of adjustment: tag-frequency controls and continuous scores.

\noindent\textbf{Tag-Frequency Control.}
A tag-frequency rule limits tagged items within a sliding window of consecutive slate positions. The indicator $\chi_r(c)\in\{0,1\}$ records whether candidate $c$ carries the tag associated with rule $r$. The permitted count lies between $T_{\min}(r)$ and $T_{\max}(r)$. At step $t$, $n_{r,t}(c)$ denotes the count after appending candidate $c$, and $\Delta_{r,t}(c)$ denotes the number of tagged items that can still be placed in the remaining window positions. Candidate $c$ violates rule $r$ in either of the following cases:

\begin{equation}
v_{r,t}(c)=
\begin{cases}
  1 & \chi_r(c)=1,\ n_{r,t}(c) > T_{\max}(r),\\
  1 & \chi_r(c)=0,\ n_{r,t}(c)+\Delta_{r,t}(c) < T_{\min}(r),\\
  0 & \text{otherwise},
\end{cases}
\label{eq:punish}
\end{equation}

The finite adjustment is $f_{r,t}(c)=-\eta_r\,v_{r,t}(c)$, with $\eta_r=0.1$. The first case penalizes a tagged candidate that would exceed the upper bound, while the second penalizes an untagged candidate when selecting it would make the minimum quota unreachable. Finite adjustments provide local control over spacing, deduplication, and quotas. Mandatory rules instead assign a $-\infty$ mask to candidates flagged by the corresponding check.

Each rule is checked independently, so passing every local check does not guarantee that a prefix has a completion satisfying all interacting quotas. Equation~\eqref{eq:decoder-policy} requires at least one candidate to remain unmasked. The aggregate violation rates in Table~\ref{tab:rule} include violations of soft business preferences and requests with infeasible candidate pools; mandatory masks therefore do not necessarily yield zero aggregate violations.

\noindent\textbf{Continuous Scores.}
A similarity penalty promotes multimodal diversity. Let $\overline{\cos}_t(c)$ denote the mean cosine similarity between candidate $c$ and the selected prefix, computed from their multimodal embeddings. The prefix is empty at $t=1$, so the similarity is defined as $0$. The adjustment $f_{\mathrm{sim},t}(c)=-\lambda_{\mathrm{sim}}\,\overline{\cos}_t(c)$ lowers the logits of redundant candidates, while $\lambda_{\mathrm{sim}}\geq 0$ controls the accuracy--diversity trade-off.

Advertising candidates receive a logit boost, with $f_{\mathrm{ad}}(c)=\beta$ for advertising candidates and $0$ otherwise. Increasing $\beta$ trades user engagement for advertiser value (Section~\ref{sec:experiments}).

\subsubsection{Reward for Business Preferences}
\label{sec:reward-combine}

For each soft business preference, we count violations across the slate using Eq.~\eqref{eq:punish}. A lower violation count yields a higher satisfaction score. We aggregate the satisfaction scores and normalize the result to $[0,1]$, obtaining the reward for business preferences $r^{\mathrm{rule}}$. The reward for business preferences is distinct from the normalized scalar reward $r^{\mathrm{RM}}$ defined in Section~\ref{sec:posttrain}.

We compare two methods for combining the rewards. \textbf{Direct Combination} adds the reward for business preferences to the scalar reward before group standardization. \textbf{Separate Standardization} standardizes the two rewards independently within each rollout group and then combines the resulting advantages. For request $q$ and rollout $g$:

\begin{equation}
A_{q,g}=A^{\mathrm{RM}}_{q,g}
+\lambda_{\mathrm{rule}}A^{\mathrm{rule}}_{q,g}.
\label{eq:rulesep}
\end{equation}

Separate Standardization places the two rewards on comparable scales, while $\lambda_{\mathrm{rule}}$ controls the contribution of the advantage derived from business preferences. Reward weighting provides only coarse control over slate composition and cannot enforce position-specific requirements. The full OneTrans-GSR system therefore combines Separate Standardization for soft preferences with constrained decoding for local business controls. Section~\ref{sec:reward-rl} compares the reward-based approaches with a Constrained Decoding Only variant that omits the reward for business preferences and handles business requirements through decoding interventions.
\subsection{Post-Training}
\label{sec:posttrain}

Geometric-Mean Reward Policy Optimization (GMR-PO) constructs a normalized scalar RM reward and standardizes it within each rollout group. The resulting advantage is combined with a separately standardized advantage derived from soft business preferences. GMR-PO then updates the decoder using fresh on-policy rollout groups with KL regularization. Algorithm~\ref{alg:gmrpo} summarizes the procedure. Each group contains $G_{\mathrm{train}}$ sampled slates for the same request (Section~\ref{sec:decoder-gen}). The logged exposure slate may be scored for monitoring but is excluded from the group statistics and policy loss.

\begin{algorithm}[t]
  \caption{OneTrans-GSR GMR-PO post-training loop}
  \label{alg:gmrpo}

  \begin{algorithmic}[1]
    \Require decoder $\pi_\theta$; frozen efficiency-RM ensemble and LTV RM
    \Require frozen reference decoder $\pi_{\mathrm{ref}}$; clipping width $\epsilon_{\mathrm{clip}}$; KL coefficient $\lambda_{\mathrm{KL}}$
    \Require rollout count $G_{\mathrm{train}}$; request-specific business rules $\rules_q$
    \Require disagreement-penalty coefficient $\kappa=0.01$; weight for business preferences $\lambda_{\mathrm{rule}}=0.1$
    \Require objective weights $w_v$ as defined in Eq.~\eqref{eq:gmrpo}

    \For{each request batch $\mathcal{Q}$}
      \State Freeze the rollout policy: $\pi_{\theta_{\mathrm{old}}}\gets\pi_\theta$
      \ForAll{$q\in\mathcal{Q}$}
        \State Sample $L_{q,1:G_{\mathrm{train}}}$ from $\pi_{\theta_{\mathrm{old}}}$ by constrained decoding using $(u_q,\cand_q,\rules_q)$ \hfill\mbox{\normalfont{Eq.~\eqref{eq:decoder-policy}}}
        \State Score each $L_{q,g}$ with the efficiency-RM ensemble and the LTV RM
        \State Apply the disagreement penalty for each objective $v\in\{\mathrm{click},\mathrm{order},\mathrm{gmv}\}$:
        \Statex \hskip\algorithmicindent\hskip\algorithmicindent\hskip\algorithmicindent $R^v_{q,g}\gets\mu^v_{q,g}-\kappa\sigma^v_{q,g}$; keep $R^{\mathrm{ltv}}_{q,g}$ unchanged
        \hfill\mbox{\normalfont{Eq.~\eqref{eq:rm-ensemble}}}
        \State Normalize each objective across $g=1{:}G_{\mathrm{train}}$ and form $r^{\mathrm{RM}}_{q,g}$ \hfill\mbox{\normalfont{Eqs.~\eqref{eq:rewardnorm} and~\eqref{eq:gmrpo}}}
        \State Compute $r^{\mathrm{rule}}_{q,1:G_{\mathrm{train}}}$, a reward encoding the soft business preferences in $\rules_q$
        \State Standardize both rewards within the group: $A^{\mathrm{RM}}_{q,g}\gets\hat{r}^{\mathrm{RM}}_{q,g}$ and $A^{\mathrm{rule}}_{q,g}\gets\hat{r}^{\mathrm{rule}}_{q,g}$
        \hfill\mbox{\normalfont{Eq.~\eqref{eq:group-znorm}}}
        \State Set $A_{q,g}\gets A^{\mathrm{RM}}_{q,g}+\lambda_{\mathrm{rule}}A^{\mathrm{rule}}_{q,g}$ and treat it as constant during the update
        \hfill\mbox{\normalfont{Eq.~\eqref{eq:rulesep}}}
      \EndFor
      \State Compute current, rollout, and reference policy probabilities at all sampled prefixes
      \State Compute the probability ratios and conditional KL divergences
      \State Compute the clipped batch loss with KL regularization and fixed advantages \hfill\mbox{\normalfont{Eq.~\eqref{eq:gmrpo-loss}}}
      \State Update $\theta$ with one optimizer step and discard the rollouts
    \EndFor
  \end{algorithmic}
  \end{algorithm}

The loss averages over the sampled slates, whereas serving selects a slate by sequence likelihood. An increase in the average sampled reward therefore does not necessarily imply an improvement under the serving-time selection rule.

The RM reward $r^{\mathrm{RM}}_{q,g}$ combines the three efficiency objectives with the LTV objective through ensemble scoring, per-objective normalization, and geometric aggregation. The resulting reward is a proxy optimization signal rather than a calibrated estimate of the expected business outcome.

\subsubsection{Ensemble Scoring}
\label{sec:hacking}

An RM trained on finite logged exposures is a noisy proxy for the underlying objective~\cite{gao2023overoptimization,skalse2022rewardhacking}. Optimizing this proxy can favor slates whose scores are inflated by estimation error. We therefore score each rollout using an ensemble with a disagreement penalty.

The efficiency RMs use different initializations. Training batches are assigned to the ensemble members in round-robin order, so each member observes a different part of the data stream. The resulting diversity is intended to reduce correlated prediction errors~\cite{coste2024ensembles,eisenstein2023herding}. For objective $v\in\mathcal{V}$, let $\mu^v_{q,g}$ and $\sigma^v_{q,g}$ denote the mean and standard deviation of the ensemble scores for request $q$ and rollout $g$. We subtract a disagreement penalty from the mean:

\begin{equation}
R^v_{q,g}=\mu^v_{q,g}-\kappa\,\sigma^v_{q,g},
\qquad \kappa=0.01.
\label{eq:rm-ensemble}
\end{equation}

The coefficient $\kappa$ controls the penalty strength. The ensemble mean combines the member predictions, while the penalty discounts scores with greater disagreement. For brevity, the experimental tables refer to Eq.~\eqref{eq:rm-ensemble} as a lower confidence bound (LCB). The score is a conservative heuristic rather than a calibrated confidence bound and provides no statistical coverage guarantee. Bias shared by all ensemble members remains uncorrected. The LTV score from Eq.~\eqref{eq:ltv} is used without an ensemble penalty.

The comparison between the single RM and ensemble variants evaluates the complete scoring pipelines and does not isolate the effect of the disagreement penalty (Section~\ref{sec:reward-rl}). The difference between offline reward gains and realized online lift is therefore used only as a deployment diagnostic, not as a direct measure of reward over-optimization.

\subsubsection{Reward Aggregation}
\label{sec:efficiency}
\noindent\textbf{Per-Objective Normalization.}
\label{sec:rewardnorm} The objectives differ in scale, and their score ranges can vary across rollout groups. We min--max normalize each objective within a group so that the sampled slates are compared on a common
scale, then apply a small positive floor:

\begin{equation}
\begin{aligned}
\bar R^v_{q,g} &=
\frac{R^v_{q,g}-\min_{j} R^v_{q,j}}
     {\max_{j} R^v_{q,j}-\min_{j} R^v_{q,j}+\epsilon_{\mathrm{mm}}},\\
\tilde R^v_{q,g} &= \max\!\left\{\delta_R,\;\bar R^v_{q,g}\right\},
\end{aligned}
\label{eq:rewardnorm}
\end{equation}

The extrema are computed over the $G_{\mathrm{train}}$ slates sampled for the same request. We use $\epsilon_{\mathrm{mm}}=10^{-8}$ and $\delta_R=10^{-4}$ and compute the statistics in FP32, as in Eq.~\eqref{eq:group-znorm}. If an objective is constant within the group, all its normalized scores equal $\delta_R$, so the objective introduces no preference among the sampled slates. The scores passed to the aggregation are therefore strictly positive.

\noindent\textbf{Geometric Aggregation.}
\label{sec:multiobj} We combine the normalized scores into a single reward using a weighted geometric mean:

\begin{equation}
r^{\mathrm{RM}}_{q,g} = \prod_{v} \big(\tilde R^v_{q,g}\big)^{w_v},
\label{eq:gmrpo}
\end{equation}

where $v\in\{\mathrm{click},\mathrm{order},\mathrm{gmv},\mathrm{ltv}\}$. The positive weights $w_v$ sum to one and are tuned as hyperparameters. GMR in GMR-PO refers to this geometric aggregation.

The geometric mean makes the combined reward sensitive to a low score on any objective, whereas an additive sum allows high scores on other objectives to offset that score. The floor $\delta_R$ prevents a low score from setting the combined reward to zero. GMR-PO applies the geometric mean across reward objectives, whereas GSPO and GMPO apply it across token-level terms in the policy objective~\cite{zheng2025gspo,zhao2025gmpo}. Geometric aggregation does not guarantee improvement on every objective or Pareto optimality. We standardize the aggregated reward using Eq.~\eqref{eq:group-znorm} to obtain $A^{\mathrm{RM}}_{q,g}$ for \textbf{Separate Standardization} in Eq.~\eqref{eq:rulesep}.

\subsubsection{Policy Update}

For each request $q$, we standardize a group signal $x_{q,1:G_{\mathrm{train}}}$ over its rollout group:

\begin{equation}
\hat{x}_{q,g}
=\frac{x_{q,g}-\operatorname{mean}(x_{q,1:G_{\mathrm{train}}})}
      {\sqrt{\operatorname{var}(x_{q,1:G_{\mathrm{train}}})+\epsilon_z}},
\label{eq:group-znorm}
\end{equation}

where the mean and population variance are computed over the $G_{\mathrm{train}}$ sampled slates. We use $\epsilon_z=10^{-8}$ and accumulate the moments in FP32. Applying Eq.~\eqref{eq:group-znorm} to $r^{\mathrm{RM}}$ and $r^{\mathrm{rule}}$ gives $A^{\mathrm{RM}}_{q,g}=\hat{r}^{\mathrm{RM}}_{q,g}$ and $A^{\mathrm{rule}}_{q,g}=\hat{r}^{\mathrm{rule}}_{q,g}$, respectively. \textbf{Separate Standardization} combines the two advantages using Eq.~\eqref{eq:rulesep}.

Let $\mathcal{Q}$ denote the current batch of recommendation requests. For request $q$, rollout $g$, and position $t$, define the policy state as $s_{q,g,t}=(c_{q,g,<t},u_q,\cand_q)$. Following GRPO~\cite{shao2024grpo}, we optimize the clipped batch loss

\begin{equation}
\begin{alignedat}{2}
\mathcal{L}_{\mathrm{GMR\text{-}PO}}(\theta)
&= -\mathbb{E}_{q\sim\mathcal{Q}}\Biggl[
\frac{1}{G_{\mathrm{train}}T}\sum_{g=1}^{G_{\mathrm{train}}}\sum_{t=1}^{T}
\Biggl\{&&
\min\Biggl[
\frac{\pi_\theta(c_{q,g,t}\mid s_{q,g,t})}
   {\pi_{\theta_{\mathrm{old}}}(c_{q,g,t}\mid s_{q,g,t})} A_{q,g},\\[4pt]
& &&\operatorname{clip}\!\left(
\frac{\pi_\theta(c_{q,g,t}\mid s_{q,g,t})}
   {\pi_{\theta_{\mathrm{old}}}(c_{q,g,t}\mid s_{q,g,t})},
1-\epsilon_{\mathrm{clip}},1+\epsilon_{\mathrm{clip}}\right) A_{q,g}
\Biggr]
\\[4pt]
& &&
-\lambda_{\mathrm{KL}}D_{\mathrm{KL}}\!\left[
\pi_\theta(\cdot\mid s_{q,g,t})\,\middle\|\,
\pi_{\mathrm{ref}}(\cdot\mid s_{q,g,t})\right]
\Biggr\}\Biggr].
\end{alignedat}
\label{eq:gmrpo-loss}
\end{equation}

Here $\pi_{\theta_{\mathrm{old}}}$ is the frozen policy used to generate the current rollout group, while $\pi_{\mathrm{ref}}$ is the frozen decoder before post-training. The parameter $\epsilon_{\mathrm{clip}}$ defines the clipping interval, and $\lambda_{\mathrm{KL}}$ controls the KL penalty that limits deviation from the reference decoder. The advantages and both frozen policies are held fixed during back-propagation. In the deployed configuration described in Algorithm~\ref{alg:gmrpo}, each fresh on-policy rollout batch is used for one optimizer step before the next batch is sampled.

\section{Experiments}
\label{sec:experiments}
\label{sec:ablation}

The experiments examine whether a single autoregressive decoder trained offline with slate-level rewards can replace the online Generator--Evaluator cascade for both reranking and mix-ranking. \textbf{RQ1} evaluates the RM-estimated utility of slates generated without an online evaluator or a downstream reranking or mix-ranking module. \textbf{RQ2} assesses the RM's predictive performance on logged exposure slates and its sensitivity to slate order and composition. \textbf{RQ3} examines slate properties associated with user experience. \textbf{RQ4} studies multi-objective optimization, soft business preferences, and controls applied during decoding. \textbf{RQ5} examines compute scaling and serving efficiency. Finally, \textbf{RQ6} compares OneTrans-GSR with the deployed Generator--Evaluator baseline in a production A/B test.

\subsection{Experimental Setup}
\label{sec:setup}

\textbf{Production Setting.} We evaluate OneTrans-GSR in a large-scale industrial recommendation system for e-commerce. The system serves product cards, short videos, livestreams, and advertising cards supplied by distinct upstream business channels. Given the candidates scored by upstream fine-ranking, the policy selects and orders a fixed-length exposure slate of distinct candidates, jointly handling reranking and content-type mixing under business constraints. We evaluate slate-level click, order, and GMV objectives, slate composition, and realized business outcomes.

\begin{table}[htbp]
  \centering
  \caption{Offline evaluation of decoder policies under the chronological next-batch protocol. The full OneTrans-GSR reference policy combines Separate Standardization with constrained decoding. Logged exposure slates from the production Generator--Evaluator (G-E) system serve as the reference. Relative gain (RG) is reported as a ratio for the three efficiency objectives and averaged over three evaluation days ($\pm$ daily standard deviation).}
  \label{tab:offline-ge}
  \small
  \setlength{\tabcolsep}{4pt}
  \begin{tabular*}{\textwidth}{@{\extracolsep{\fill}}llccc@{}}
    \toprule
    & & \multicolumn{3}{c}{Decoder Policy RG vs. Logged G-E} \\
    \cmidrule(l){3-5}
    Type & Variant & ListClick$\uparrow$ & ListOrder$\uparrow$ & ListGMV$\uparrow$ \\
    \midrule
    Reference Policy & OneTrans-GSR & 0.0768{\scriptsize$\pm$0.0275} & 0.3132{\scriptsize$\pm$0.0581} & 0.2870{\scriptsize$\pm$0.0590} \\
    \midrule
    \multirow{2}{*}{Cand. Interaction}
      & Output-Only Interaction     & 0.0420{\scriptsize$\pm$0.0280} & 0.2710{\scriptsize$\pm$0.0280} & 0.2070{\scriptsize$\pm$0.0390} \\
      & Cross-Attention Interaction & 0.0640{\scriptsize$\pm$0.0240} & 0.2710{\scriptsize$\pm$0.0780} & 0.2480{\scriptsize$\pm$0.0740} \\
    \midrule
    \multirow{3}{*}{Decoder}
      & w/o Autoregressive Decoding & $-0.2413${\scriptsize$\pm$0.1246} & $-0.4530${\scriptsize$\pm$0.1301} & $-0.5426${\scriptsize$\pm$0.1226} \\
      & w/o Stop-Gradient           & 0.0540{\scriptsize$\pm$0.0210} & 0.2050{\scriptsize$\pm$0.0480} & 0.2010{\scriptsize$\pm$0.0520} \\
      & w/o Fine-Ranking Score      & $-0.0033${\scriptsize$\pm$0.0009} & $-0.0048${\scriptsize$\pm$0.0014} & $-0.0059${\scriptsize$\pm$0.0011} \\
    \midrule
    \multirow{2}{*}{Reward}
    & w/o Content-Type Preference & 0.0830{\scriptsize$\pm$0.0210} & 0.3300{\scriptsize$\pm$0.0420} & 0.2580{\scriptsize$\pm$0.0320} \\
    & w/o LTV & 0.0607{\scriptsize$\pm$0.0275} & 0.3403{\scriptsize$\pm$0.0581} & 0.2748{\scriptsize$\pm$0.0590} \\
    \midrule
    \multirow{4}{*}{RL Policy}
      & w/o RL Post-Training & 0.0676{\scriptsize$\pm$0.0027} & 0.0119{\scriptsize$\pm$0.0049} & $-0.0400${\scriptsize$\pm$0.0055} \\
      & GRPO                 & 0.0748{\scriptsize$\pm$0.0256} & 0.2610{\scriptsize$\pm$0.0348} & 0.2060{\scriptsize$\pm$0.0620} \\
      & GDPO                 & 0.1488{\scriptsize$\pm$0.0114} & 0.2515{\scriptsize$\pm$0.0753} & 0.1050{\scriptsize$\pm$0.1050} \\
      & HERO                 & 0.0770{\scriptsize$\pm$0.0278} & 0.2485{\scriptsize$\pm$0.0548} & 0.1310{\scriptsize$\pm$0.0910} \\
    \bottomrule
  \end{tabular*}
\end{table}

\textbf{Dataset.} We use three months of proprietary production logs from this system. Each request records the candidate pool available to reranking and mix-ranking, the displayed exposure slate, and the corresponding outcome labels. The displayed slates provide supervision for RM training, while the retained candidate pools allow the policies to generate alternative slates over the logged candidates. Outcome labels are available only for the displayed slates; unexposed candidates and counterfactual slates have no observed outcomes in these logs.

\textbf{Baselines.} The production Generator--Evaluator (G-E) system~\cite{feng2021prs,shi2023pier} serves as the baseline throughout the evaluation. Multiple specialized generators propose candidate slates, and a separately trained evaluator selects one for serving. The resulting exposure logs provide the reference slates for offline decoder evaluation, while the deployed G-E system serves as the control in the production A/B test.

\textbf{Evaluation Protocol.} Offline comparisons follow the chronological next-batch protocol of OneTrans~\cite{zhang2025onetrans}: data are processed in chronological order, and each mini-batch is evaluated before being used for training. Following Section~\ref{sec:decoder}, each variant \(m\) samples \(G_{\mathrm{serve}}\) candidate slates for request \(q\) and selects the slate with the highest accumulated sequence log-probability, denoted by \(\hat{L}_q^m\). All variants use the same offline sampling budget and selection procedure. Decoder evaluation considers the selected slate rather than the average over the post-training rollout group.

Decoder evaluation compares the RM score of each selected slate with that of the corresponding logged exposure slate. The analysis considers these proxy gains alongside RM performance on logged slates and sensitivity to slate order and composition. Category diversity, multimodal similarity, fatigue, and discovery characterize the composition of the generated slates. The production A/B test separately evaluates realized user and business outcomes.

\emph{Decoder Policy Utility.} We evaluate each selected slate using RM-estimated relative gain (RG), with the logged Generator–Evaluator exposure slate as the reference. RG is reported for the three efficiency objectives: ListClick, ListOrder, and ListGMV. Let \(L_q^{\mathrm{ref}}\) denote the logged exposure slate for request \(q\), and let \(\mathcal{Q}\) denote the evaluation request set. For objective \(v\), the relative gain of policy variant \(m\) is

\begin{equation}
\operatorname{RG}_{v}(m;\mathrm{ref})
=
\frac{
\mathbb{E}_{q\in\mathcal{Q}}
  \left[r_v(\hat{L}_q^m,q)\right]
-
\mathbb{E}_{q\in\mathcal{Q}}
  \left[r_v(L_q^{\mathrm{ref}},q)\right]
}{
\left|
\mathbb{E}_{q\in\mathcal{Q}}
  \left[r_v(L_q^{\mathrm{ref}},q)\right]
\right|
+\epsilon
},
\label{eq:offline-rg}
\end{equation}

where \(r_v(L,q)\) is the objective-specific RM score and \(\epsilon>0\) is a numerical stabilizer. RG is reported as a ratio rather than a percentage. The metric quantifies a relative difference between RM scores, not a calibrated change in expected clicks, orders, or GMV. Section~\ref{sec:online} reports the production A/B results separately.

\emph{Reward Model Quality.} We evaluate the RM on logged exposure slates through predictive discrimination, aggregate click agreement, and sensitivity to slate order and composition. Slate-level AUC measures discrimination of the positive-feedback indicator \(o_v(L)\) for ListClick, ListOrder, and ListGMV; it does not measure errors in nonzero event counts or GMV amounts. Pointwise AUC is computed for EXPO, CTR, CVR, and GMV and averaged across exposure positions. For ListClick, we evaluate aggregate calibration using the predicted-to-observed click ratio (PCOC), defined as the total predicted click count divided by the total observed click count. We report PCOC separately for the direct listwise prediction and the sum of the pointwise predictions, with values closer to \(1.0\) indicating better aggregate calibration. PCOC does not assess calibration within individual user or content-type groups, nor does it imply that the post-training reward \(R^v=\exp(z^v)\) is calibrated. Controlled perturbations measure the RM's sensitivity to slate order and composition.

\emph{Slate Quality Proxies.} We use category diversity, multimodal similarity, fatigue, and discovery as proxies for slate composition and exposure quality. Category diversity counts the distinct categories in a slate. Multimodal similarity is the mean pairwise cosine similarity among the item embeddings. Fatigue measures repeated exposure to categories without observed clicks, while discovery measures the proportion of novel items. Higher category diversity and discovery and lower similarity and fatigue are preferred under these definitions. Table~\ref{tab:slate-experience} reports signed relative changes from each baseline; positive and negative values indicate increases and decreases in the underlying metric, respectively. The four metrics characterize generated slates but do not directly measure user responses.

\textbf{Hyperparameters.} The efficiency RM is an ensemble of three independently initialized models, each comprising two OneTrans layers with a hidden size of 512 and four attention heads. The decoder comprises four OneTrans layers with a hidden size of 256 and four attention heads. Unless otherwise specified, all offline experiments use the three-month dataset described above. Post-training uses a global request batch size of 1,024 requests, a learning rate of \(10^{-4}\), \(\lambda_{\mathrm{rule}}=0.1\), \(\eta_r=0.1\), and \(\kappa=0.01\), with all trainable parameters optimized using AdamW. The KL coefficient \(\lambda_{\mathrm{KL}}\) is a tunable hyperparameter of the policy update. The clipping width \(\epsilon_{\mathrm{clip}}\) parameterizes the general surrogate but does not affect the deployed single-step update. Unless otherwise specified, we set \(G_{\mathrm{train}}=G_{\mathrm{serve}}=32\). Post-training samples \(G_{\mathrm{train}}\) slates per request to compute group advantages. Offline evaluation and serving sample \(G_{\mathrm{serve}}\) slates and select the one with the highest accumulated sequence log-probability. All offline variants use the same sampling budget and selection procedure.

\subsection{RQ1: End-to-End Generation with a Single Decoder}
\label{sec:decoder-design}

RQ1 examines whether a single decoder can generate slates with higher RM-estimated utility from a request-specific candidate pool. As shown in Table~\ref{tab:offline-ge}, the full OneTrans-GSR model achieves positive RM-estimated RG on ListClick, ListOrder, and ListGMV relative to the logged G-E exposure slates. During evaluation, the decoder selects a generated slate solely by its accumulated sequence log-probability, without an online RM or a separate downstream reranking or mix-ranking module. Sections~\ref{sec:slate-experience} and~\ref{sec:online} separately evaluate slate properties associated with user experience and realized production outcomes.

\noindent\textbf{Autoregressive Decoding.} The non-autoregressive variant predicts all slate positions in parallel while retaining the same inputs, RM, business rules, and training configuration. Removing autoregressive decoding makes the RG negative for all three objectives, with the largest degradation observed on ListGMV. The comparison therefore favors prefix-conditioned generation over the evaluated parallel decoding baseline.

\noindent\textbf{RL Post-Training.} The variant without RL post-training retains decoder pretraining and supervised adaptation but omits optimization against the RM. Removing post-training has a limited effect on ListClick but substantially reduces ListOrder and ListGMV, with the latter falling below the logged G-E reference. The RM-score comparisons therefore indicate that post-training contributes most strongly to the conversion objectives.

\noindent\textbf{Candidate Interaction.} We compare the output-only, cross-attention, and prefill designs described in Section~\ref{sec:decoder-prefill}, which introduce candidate interaction at progressively earlier stages of the decoder. Output-only scoring defers candidate comparison to the final dot-product head, cross-attention introduces candidates within each decoder layer, and prefill places them directly in the input context. Earlier interaction consistently improves RG, with prefill achieving the strongest result on all three objectives. Prefill also avoids the per-layer overhead introduced by cross-attention and is therefore used as the default design.

\noindent\textbf{Gradient Isolation and Input Prior.} Removing the output-side stop-gradient from the candidate embeddings lowers RG on all three objectives, consistent with the gradient conflict discussed in Section~\ref{sec:decoder-sg}. Removing the upstream OneTrans fine-ranking score has a larger effect and makes all three RG values negative. The two ablations support output-side gradient isolation and the use of the fine-ranking score as a pointwise prior for slate generation.

\subsection{RQ2: Slate-Level Reward Modeling}
\label{sec:listwise-interaction}
RQ2 evaluates the predictive quality and design of the RM. We compare its discrimination with representative ranking architectures, examine slate-level supervision and readout design through AUC and aggregate calibration, probe its sensitivity to changes in slate order and composition, and ablate the main RM components.

\noindent\textbf{Pointwise Predictive Quality.} We first evaluate the RM on CTR and CVR, the pointwise objectives shared with the upstream fine-ranker. Table~\ref{tab:pointwise-auc} shows positive relative AUC gains on both objectives, with a larger gain on CTR. The results indicate improved item-level discrimination of the observed click and conversion outcomes under the logged-data protocol.

\begin{table}[t]
\centering
\caption{Relative pointwise AUC gains (\%) of the OneTrans-GSR RM over the upstream fine-ranker on their shared CTR and CVR objectives. AUC is averaged across exposure positions.}
\label{tab:pointwise-auc}
\small
\begin{tabular}{lc}
  \toprule
  Target & Relative AUC Gain (\%) $\uparrow$ \\
  \midrule
  CTR & \(+1.23\) \\
  CVR & \(+0.62\) \\
  \bottomrule
\end{tabular}
\end{table}

\noindent\textbf{Slate-Level Predictive Comparison.} We next compare the OneTrans-GSR RM with two representative slate-level ranking methods. RankFormer~\cite{buyl2023rankformer} uses bidirectional self-attention for context-aware item scoring and a CLS-based head for listwide quality prediction, whereas TSPRank~\cite{li2025tsprank} models ranking as a path over pairwise item transitions. The OneTrans-GSR RM scores the complete slate in exposure order using causal attention and a terminal EOS readout. Causal self-attention conditions each item representation on the preceding items, and a terminal EOS token summarizes the sequence for slate-level prediction. Auxiliary pointwise heads provide item-level supervision, allowing the RM to learn item outcomes and slate utility jointly. All models use the same dataset, input features, and chronological next-batch protocol.

Table~\ref{tab:listwise-comparison} reports the largest AUC point estimates for the OneTrans-GSR RM on both ListClick and ListOrder, at \(0.8162\) and \(0.8976\), respectively. RankFormer nearly matches its ListOrder AUC at \(0.8973\) and reports ListClick AUC of \(0.8106\); TSPRank reports \(0.7254\) and \(0.7548\). The clearest observed separation is on ListClick. These comparisons concern logged-data discrimination under the stated protocol. The following ablations examine supervision and readout design using AUC, aggregate PCOC, and structural sensitivity.

\begin{table}[t]
\centering
\caption{Listwise predictive discrimination of representative listwise ranking models under the same dataset, input features, and chronological next-batch evaluation
protocol. Higher values are better.}
\label{tab:listwise-comparison}
\small
\begin{tabular}{lcc}
  \toprule
  Model
  & ListClick AUC $\uparrow$
  & ListOrder AUC $\uparrow$ \\
  \midrule
  RankFormer~\cite{buyl2023rankformer}
  & 0.8106
  & 0.8973 \\
  TSPRank~\cite{li2025tsprank}
  & 0.7254
  & 0.7548 \\
  OneTrans-GSR RM
  & \textbf{0.8162}
  & \textbf{0.8976} \\
  \bottomrule
\end{tabular}
\end{table}

\noindent\textbf{Slate-Level Supervision and Readout.} We compare supervision and readout design using pointwise and slate-level discrimination and aggregate calibration. The slate-level variants differ only in how they construct a representation of the slate: pooling aggregates the item representations, concatenation joins the representations across positions, and the EOS readout uses the final state of a terminal token. Pooling and concatenation improve ListClick discrimination but reduce pointwise CTR discrimination relative to the pointwise baseline. In contrast, the EOS readout improves both AUC measures and produces the PCOC value closest to \(1.0\). The combined discrimination and calibration results support the EOS readout for the evaluated RM.

\begin{table*}[t]
\centering
\caption{Effect of slate-level supervision and aggregation method on pointwise CTR and listwise ListClick discrimination, and aggregate predicted-to-observed click agreement. Higher AUC is better; PCOC values closer to \(1.0\) indicate closer aggregate agreement.}
\label{tab:listwise-aggregation}
\small
\begin{tabular}{llccc}
  \toprule
  Supervision
  & Aggregation Method
  & Pointwise CTR AUC $\uparrow$
  & Listwise ListClick AUC $\uparrow$
  & PCOC $\rightarrow 1$ \\
  \midrule
  Pointwise
  & Sum of Pointwise Scores
  & 0.8119
  & 0.8154
  & 1.09 \\
  \midrule
  \multirow{3}{*}{Slate-Level}
  & Pooling
  & 0.8109
  & 0.8158
  & 0.98 \\
  & Concatenation
  & 0.8096
  & 0.8160
  & 1.06 \\
  & EOS Token
  & \textbf{0.8120}
  & \textbf{0.8162}
  & \textbf{1.00} \\
  \bottomrule
\end{tabular}
\end{table*}

\noindent\textbf{Sensitivity to Slate Structure.} Table~\ref{tab:list-stress} compares the responses of three configurations to controlled changes in slate order and composition. The
configurations differ in their training objective, either pointwise or slate-level, and their readout, either summed predictions or the EOS state. All score differences use the same scale.
Because the perturbed slates have no outcome labels, the experiment measures sensitivity rather than predictive accuracy.

\begin{table}[t]
\centering
\caption{Sensitivity of the ListGMV score to controlled slate perturbations. Values are score differences between the perturbed and original slates under position swaps and nonadjacent same-category replacements. All configurations use the same score scale; larger-magnitude changes indicate greater sensitivity to slate order and composition.}
\label{tab:list-stress}
\small
\setlength{\tabcolsep}{10pt}
\begin{tabular}{@{}llccccc@{}}
  \toprule
  & & \multicolumn{2}{c}{Reorder}
  & \multicolumn{3}{c}{Same-Category Replacements} \\
  \cmidrule(lr){3-4}
  \cmidrule(l){5-7}
  Objective & Readout
  & 1$\leftrightarrow$Mid
  & 1$\leftrightarrow$Tail
  & Replace 2
  & Replace 3
  & Replace 5 \\
  \midrule
  Pointwise
  & Summed
  & $-0.0098$
  & $-0.0104$
  & $-0.0017$
  & $-0.0017$
  & $-0.0018$ \\
  Slate-Level
  & Summed
  & $-0.0083$
  & $-0.0075$
  & $-0.0016$
  & $-0.0015$
  & $-0.0017$ \\
  Slate-Level
  & EOS Token
  & $\mathbf{-0.1053}$
  & $\mathbf{-0.1567}$
  & $\mathbf{-0.0065}$
  & $\mathbf{-0.0159}$
  & $\mathbf{-0.0208}$ \\
  \bottomrule
\end{tabular}
\end{table}

The first perturbation swaps the item at the first position with an item at a middle or tail position while preserving the item set. The summed readouts change only slightly, indicating limited sensitivity to exposure order. The EOS score responds by more than an order of magnitude more, with a larger change for the tail-position swap.

The second perturbation replaces an increasing number of nonadjacent items with candidates from the same category while preserving the slate length. The response of the EOS readout grows with the number of replacements, whereas both summed readouts remain nearly unchanged. Slate-level supervision alone therefore produces little structural sensitivity when combined with a summed readout. However, replacement also changes item identity and may alter individual relevance, so the observed score changes do not isolate a pure diversity effect or establish that the true slate utility decreases. Together with the logged-data AUC and PCOC results, these probes support the EOS readout as more sensitive to ordered slate structure in the evaluated setting.

\noindent\textbf{RM Component Ablations.}
\label{sec:rm-component-ablation} We next ablate the content-type preference component and the upstream fine-ranking score. As defined in Section~\ref{sec:rm-heads}, the auxiliary preference heads cover product, live, and video candidates. Advertising candidates remain in the candidate pool but are excluded from this auxiliary task. Table~\ref{tab:rm-quality} shows that removing either component reduces AUC across all pointwise and slate-level targets. Removing content-type preference has its largest effect on GMV discrimination, whereas removing the fine-ranking score produces broader degradation across conversion and relevance targets. The logged-data results suggest complementary roles: content-type preference provides information useful for GMV prediction, while the fine-ranking score supplies a relevance prior across tasks.

\begin{table*}[t]
\centering
\caption{Relative AUC changes from the full OneTrans-GSR RM under component ablations. Pointwise AUC is averaged across exposure positions. Negative values indicate degradation after removing the corresponding component.}
\label{tab:rm-quality}
\small
\setlength{\tabcolsep}{4pt}
\begin{tabular}{lccccccc}
  \toprule
  & \multicolumn{3}{c}{Listwise AUC Change (\%)}
  & \multicolumn{4}{c}{Pointwise AUC Change (\%)} \\
  \cmidrule(lr){2-4}
  \cmidrule(lr){5-8}
  Variant
  & ListClick
  & ListOrder
  & ListGMV
  & EXPO
  & CTR
  & CVR
  & GMV \\
  \midrule
  Full OneTrans-GSR RM
  & - & - & -
  & - & - & - & - \\
  w/o Content-Type Preference
  & -1.4\% & -3.2\%  & -6.8\% 
  & -2.1\%  & -4.2\%  & -3.5\%  & -5.9\%  \\
  w/o Fine-Ranking Score
  & -4.3\%  & -7.7\%  & -5.6\% 
  & -2.3\%  & -6.0\%  & -7.0\%  & -7.4\%  \\
  \bottomrule
\end{tabular}
\end{table*}

\subsection{RQ3: Slate Composition and Exposure Characteristics}
\label{sec:slate-experience}

RQ3 evaluates slate composition and exposure characteristics that may affect user experience, complementing the RM-estimated utility results in RQ1. Following Section~\ref{sec:setup}, we use category diversity, multimodal similarity, fatigue, and discovery as offline proxies rather than direct measures of user response.

\noindent\textbf{Evaluation Protocol.} We compare two decoding settings of OneTrans-GSR. Panel A sets the additive logit adjustment \(\delta_t(c)\) in Eq.~\eqref{eq:decoder-policy} to zero while retaining the availability mask \(m_{t,i}\), which prevents repeated selection. The comparison baselines sort the same candidate pool by CTR, CTCVR, GMV, or the production ranking score and select the top \(T\) items. Panel A also compares the generated slates with the logged G-E exposure slates. Panel B restores \(\delta_t(c)\) and uses the same logged G-E reference. The policy parameters, training objectives, requests, candidate pools, and reference slates remain unchanged, allowing the matched comparison to measure the combined effect of the decoding adjustments.

\begin{table*}[t]
\centering
\caption{Relative changes (\%) in offline slate-quality proxies for OneTrans-GSR against score-sorted and logged Generator--Evaluator (G-E) baselines. Panel A disables the additive logit adjustments, while Panel B enables them and uses the logged G-E exposure as the reference. Positive and negative values indicate increases and decreases in the underlying metric; arrows show the preferred direction. MM denotes multimodal.}
\label{tab:slate-experience}
\small
\begin{tabular*}{0.96\textwidth}{@{\extracolsep{\fill}}l@{\hspace{35pt}}rrrr@{}}
  \toprule
  Baseline
  & Category Diversity $\uparrow$
  & MM Similarity $\downarrow$
  & Fatigue $\downarrow$
  & Discovery $\uparrow$ \\
  \midrule

  \multicolumn{5}{@{}l}{\textit{Panel A: Logit Adjustments Disabled}} \\
  \addlinespace[2pt]
  CTR Descending
  & +46.88\%
  & -15.38\%
  & -40.91\%
  & +51.24\% \\
  CTCVR Descending
  & +38.03\%
  & -8.33\%
  & -35.84\%
  & +22.06\% \\
  GMV Descending
  & +35.25\%
  & -8.23\%
  & -35.00\%
  & +25.10\% \\
  Production Ranking Formula
  & +17.79\%
  & -7.20\%
  & -19.41\%
  & -6.53\% \\
  Logged G-E Exposure
  & +0.11\%
  & -0.05\%
  & +3.41\%
  & -2.47\% \\
  \midrule
  \multicolumn{5}{@{}l}{\textit{Panel B: Logit Adjustments Enabled}} \\
  \addlinespace[2pt]
  Logged G-E Exposure
  & +6.93\%
  & -5.23\%
  & -7.65\%
  & -1.48\% \\
  \bottomrule
\end{tabular*}
\end{table*}

\noindent\textbf{Comparison with Fine-Ranking Baselines.} Panel A evaluates the learned policy without additive logit adjustments from business rules. Relative to all four score-sorted baselines, OneTrans-GSR produces greater category diversity, lower multimodal similarity, and lower fatigue. Discovery also improves over the CTR, CTCVR, and GMV baselines but remains lower than that of the production ranking score. Because these patterns remain when \(\delta_t(c)=0\), they arise from the learned policy rather than from the decoding adjustments.

\noindent\textbf{Comparison with Logged Production Slates.} Without the additive logit adjustments, OneTrans-GSR closely matches the logged G-E slates in category diversity and multimodal similarity. However, the generated slates exhibit higher fatigue and lower discovery. The comparison indicates that the learned policy captures much of the composition of the production slates but does not fully reproduce their fatigue and discovery patterns.

\noindent\textbf{Effect of Logit Adjustments.} Panel B restores \(\delta_t(c)\) and compares the resulting slates with the same logged G-E reference. Enabling the adjustments improves all four proxies relative to the setting in Panel A. Category diversity increases beyond the logged baseline, while multimodal similarity and fatigue fall below it. The discovery gap also narrows but does not close completely. The decoding adjustments therefore complement the learned policy by controlling composition patterns that are not fully captured through policy training alone.

Section~\ref{sec:online} evaluates the deployed policy using realized business outcomes and end-to-end latency across the reranking and mix-ranking stages.

\subsection{RQ4: Multi-Objective Optimization and Business Control}
\label{sec:reward-rl}

RQ4 examines how OneTrans-GSR balances RM-scored objectives and business requirements. We first compare methods for multi-objective post-training and then evaluate how a separate reward for soft business preferences and decoding controls affect rule compliance. Hard masks restrict candidate selection, while finite logit adjustments provide soft steering. Steering advertiser value illustrates the latter mechanism. We finally compare post-training with a single RM and an ensemble scorer using both offline RM scores and production outcomes.

\noindent\textbf{Multi-Objective Post-Training.} Table~\ref{tab:offline-ge} compares GMR-PO with GRPO~\cite{shao2024grpo}, GDPO~\cite{gdpo2026}, and HERO~\cite{hero2025} under the same decoder, reward models, and rollout configuration. All four methods achieve positive RM-estimated RG on ListClick, ListOrder, and ListGMV relative to the logged G-E slates. GMR-PO has the highest reported mean RG on ListOrder and ListGMV, while GDPO has the highest mean ListClick RG. No method therefore dominates the reported means across all three objectives. We select GMR-PO for its stronger joint performance on ListOrder and ListGMV while retaining positive ListClick RG. The daily standard deviations alone do not establish a statistically significant ranking among the methods.

\noindent\textbf{Reward Components.} Table~\ref{tab:offline-ge} examines the auxiliary content-type preference task and the LTV reward. The preference heads provide auxiliary supervision to the shared RM backbone (Section~\ref{sec:rm-heads}); content-type preference is not an additional objective in the geometric reward aggregation. Removing the preference component improves ListClick and ListOrder RG but reduces ListGMV RG. Removing the LTV reward instead reduces ListClick and ListGMV RG while increasing ListOrder RG. Both ablations therefore shift the balance among the RM-estimated objectives rather than improving all objectives simultaneously. Section~\ref{sec:online} reports the repurchase-rate result for the complete treatment and does not isolate the contribution of the LTV reward.

\begin{figure*}[t]
  \centering
  \includegraphics[width=0.85\textwidth]{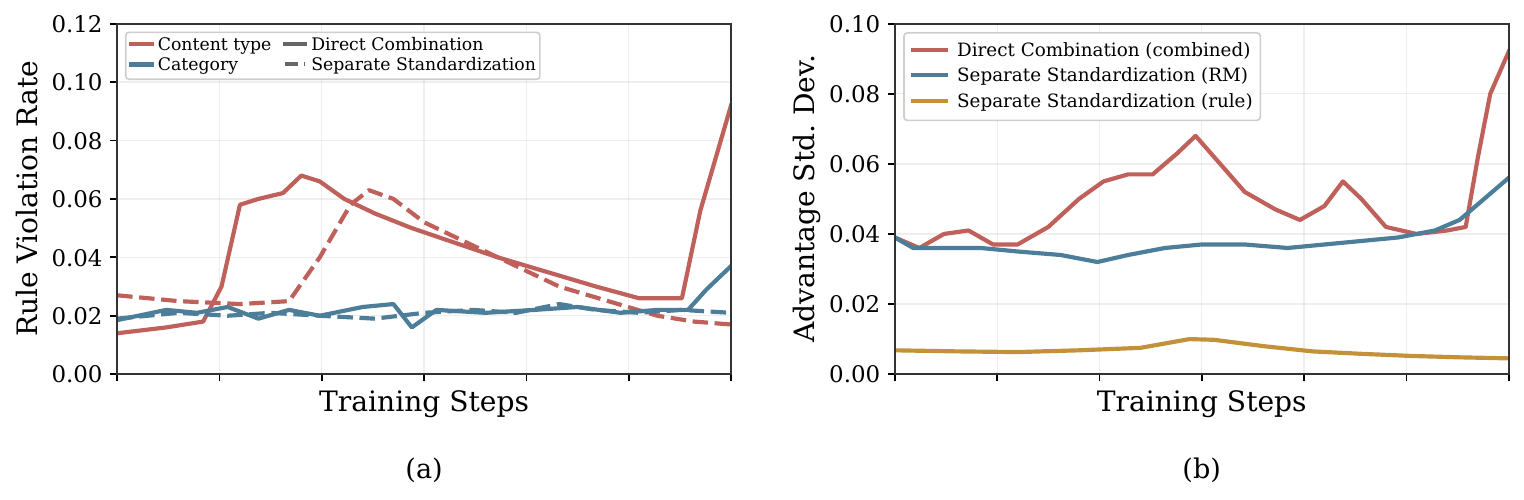}
  \caption{Optimization with rewards for business preferences under Direct Combination and Separate Standardization: (a) content-type and category-scatter violation rates; (b) advantage standard deviations.}
  \Description{Two panels comparing Direct Combination with Separate Standardization during policy post-training. Panel (a) shows content-type and category-scatter violation rates, and panel (b) shows advantage standard deviations.}
  \label{fig:rule-optimization}
  \end{figure*}

\noindent\textbf{Optimizing Soft Business Preferences.} We compare two methods for combining the scalar RM reward with the reward encoding soft business preferences (Section~\ref{sec:reward-combine}). \textbf{Direct Combination} adds the rewards before group standardization, whereas \textbf{Separate Standardization} standardizes each reward independently before combining the resulting advantages. Figure~\ref{fig:rule-optimization} shows a late-stage increase in rule violations under Direct Combination, while Separate Standardization maintains lower violation rates. The compliance results in Table~\ref{tab:rule} support separate standardization in the evaluated configuration.

\noindent\textbf{Constrained Decoding Only.} We evaluate a separate variant that retains RM-based utility optimization but omits the reward for business preferences during post-training. Business requirements are handled only through decoding interventions. This variant achieves lower reported violation rates than the reward-based approaches while preserving positive RM-estimated ListClick and ListGMV RG (Table~\ref{tab:rule}). The full OneTrans-GSR system combines Separate Standardization with constrained decoding. The aggregate violation rates include soft-preference targets and requests with infeasible candidate pools; they are not conditional hard-rule violation rates computed only over jointly feasible requests. The masks exclude candidates that violate the implemented stepwise checks, but individual quota checks do not guarantee that all interacting constraints can be satisfied when generation terminates (Section~\ref{sec:decode-intervention}).

\begin{table}[t]
    \centering
    \caption{Rule compliance and RM-estimated offline utility for alternative business-control strategies. Constrained Decoding Only retains RM-based utility optimization and handles business requirements through decoding interventions, omitting the reward for business preferences. The full system combines Separate Standardization with constrained decoding. Violation rates include soft-preference targets and requests with
    infeasible candidate pools; they are not conditional hard-rule violation rates over jointly feasible requests. Offline
    relative gain (RG) is measured against logged G-E slates and reported as the mean \(\pm\) daily standard deviation over three evaluation days. RG is omitted for Direct Combination because
    of late-stage instability.}
    \label{tab:rule}
    \small
    \begin{tabular*}{\columnwidth}{@{\extracolsep{\fill}}lcccc@{}}
      \toprule
      & \multicolumn{2}{c}{Violation Rate (\%) $\downarrow$}
      & \multicolumn{2}{c}{Offline RG $\uparrow$} \\
      \cmidrule(lr){2-3}
      \cmidrule(lr){4-5}
      Method
      & Content Type
      & Category Scatter
      & Slate Click
      & Slate GMV \\
      \midrule
      Direct Combination
      & 3.97
      & 2.10
      & \multicolumn{2}{c}{Not Reported} \\
      Separate Standardization
      & 3.08
      & 2.01
      & \textbf{\(0.0768 \pm 0.0275\)}
      & \textbf{\(0.2870 \pm 0.0590\)} \\
      Constrained Decoding Only
      & 0.11
      & 0.30
      & \(0.0698 \pm 0.0279\)
      & \(0.2660 \pm 0.0730\) \\
      \bottomrule
    \end{tabular*}
  \end{table}

\begin{figure*}[t]
\centering
\includegraphics[width=0.85\textwidth]{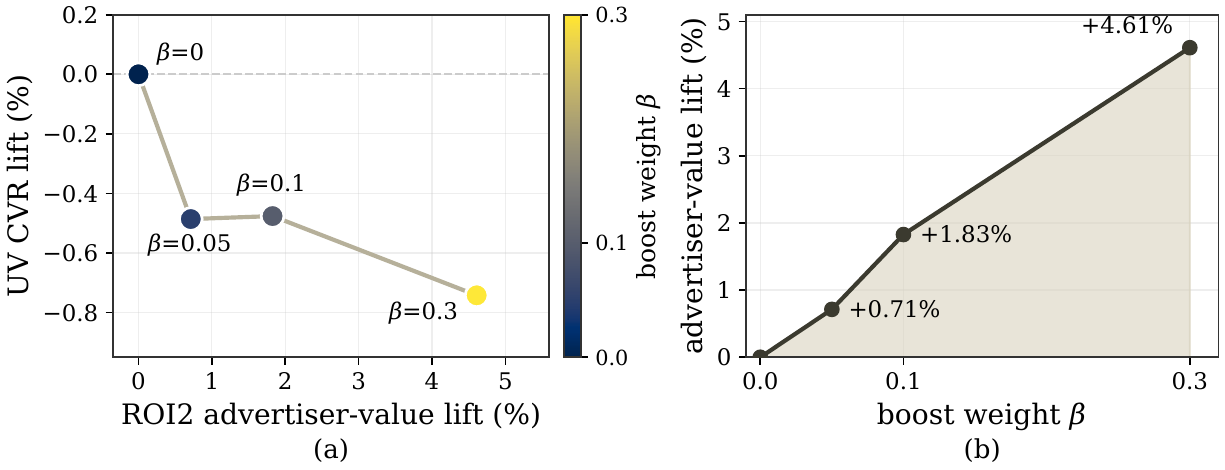}
\caption{Steering advertiser value with additive logit boosts. (a) Trade-off between advertiser-value lift and UV CVR across boost weights. (b) Advertiser-value lift as a function of the
  boost weight \(\beta\).}
\Description{Two panels showing advertiser-value steering. The first shows the trade-off between advertiser-value lift and UV CVR, and the second shows advertiser-value lift across boost weights.}
\label{fig:advertiser-steering}
\end{figure*}

\noindent\textbf{Business Steering: Advertiser Value.} We vary the constant logit boost \(\beta\) applied to advertising candidates (Section~\ref{sec:decode-intervention}). Figure~\ref{fig:advertiser-steering} shows that advertiser value increases monotonically over the tested range. Every nonzero boost also reduces user-level conversion rate (UV CVR), with the largest decline at the highest tested weight. The sweep therefore reveals a trade-off between advertiser value and UV CVR. Selecting a production operating point requires considering both outcomes alongside rule compliance.

\noindent\textbf{Reward Reliability.} Table~\ref{tab:offline-online} compares two treatment arms in the same randomized production A/B test: post-training with a single RM and post-training with the disagreement-penalized ensemble scorer referred to as LCB in Section~\ref{sec:posttrain}. Both arms share the same G-E control and differ only in the reward scorer. The single-RM arm achieves higher offline ListGMV RG but lower online GMV/u lift than the ensemble arm. The reversed ordering is consistent with errors in the proxy reward, including possible reward over-optimization, but does not identify or quantify their cause because offline ListGMV RG and online GMV/u are different endpoints. The ensemble scorer also changes both prediction averaging and the disagreement penalty, with its members trained on different data streams. The comparison therefore evaluates the complete scoring pipelines rather than isolating the disagreement penalty.

\begin{table}[t]
\centering
\caption{Offline RM-estimated ListGMV RG and online GMV/u lift for two treatment arms in the same randomized production A/B test. The arms share the G-E control and differ in the post-training reward scorer. Offline RG is averaged over three evaluation days. \(^{*}\) denotes \(p<0.05\) relative to the control.}
\label{tab:offline-online}
\small
\begin{tabular}{lcc}
  \toprule
  Scorer
  & Offline ListGMV RG (\%)
  & Online GMV/u Lift (\%) \\
  \midrule
  Single RM
  & 33.70
  & \(1.41^{*}\) \\
  Ensemble LCB (\(\kappa=0.01\))
  & 28.70
  & \(2.63^{*}\) \\
  \bottomrule
\end{tabular}
\end{table}

The experiments distinguish the mechanisms by their measured effects. GMR-PO balances the RM-scored objectives, Separate Standardization incorporates rewards for soft business preferences, and decoding controls reduce violations or steer exposure. The comparison between reward scorers provides a production diagnostic of reward reliability. Section~\ref{sec:online} reports business outcomes for the complete deployed treatment.

\subsection{RQ5: Training-Time Scaling and Serving Efficiency}
\label{sec:scaling}

RQ5 studies how RM capacity, decoder capacity, and sampling budgets affect predictive discrimination and RM-estimated policy RG. We separately evaluate cumulative system optimizations for decoder training and the serving model path.

\begin{figure}[t]
\centering
\includegraphics[width=0.75\linewidth]{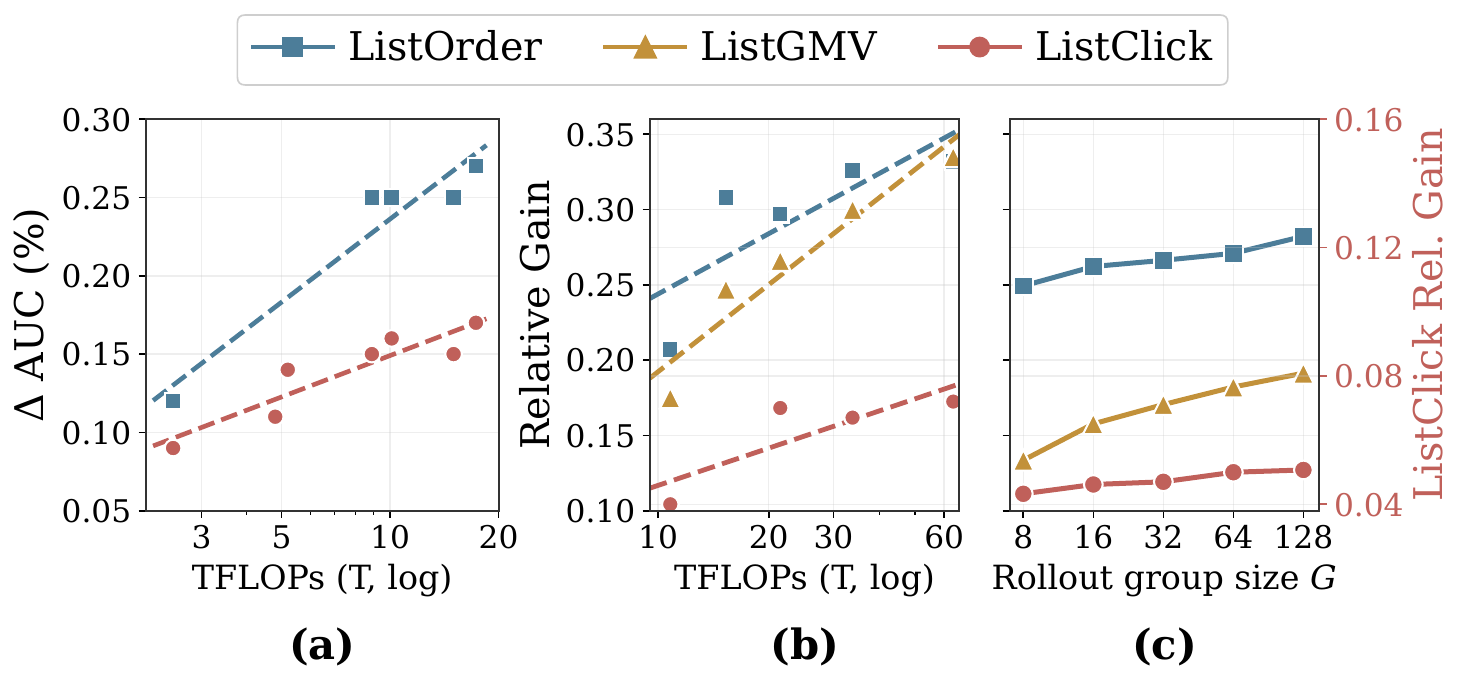}
\caption{Scaling results along three compute dimensions. (a) RM discrimination (\(\Delta\mathrm{AUC}\)) versus RM compute. (b) RM-estimated offline RG versus decoder compute. Panels (a) and (b) use logarithmic compute axes, and dashed lines show log-linear fits to the tested configurations. (c) RM-estimated offline RG versus sample count \(G\) on a logarithmic axis. Panels (b) and (c) plot ListClick against the secondary scale shown on the right.}
\label{fig:scaling}
\Description{Three panels showing reward-model discrimination as a function of RM compute, offline relative gain as a function of decoder compute, and offline relative gain as a function of the plotted sample count. The horizontal axes are logarithmic, and ListClick uses a secondary scale shown on the right of the sample-count panel.}
\end{figure}

\noindent\textbf{Compute Scaling.} Figure~\ref{fig:scaling} varies RM capacity, decoder capacity, and sample count. The capacity sweeps change the number of layers and hidden dimensions and report the resulting computation in TFLOPs. Panel~(a) shows an increase in RM \(\Delta\mathrm{AUC}\) with compute, reflecting predictive discrimination rather than realized slate utility. Panel~(b) shows positive fitted trends in ListOrder and ListGMV RG as decoder compute increases, while ListClick RG remains comparatively flat. Panel~(c) shows improvements in all three RG metrics across the tested sample counts, with diminishing marginal gains at larger values. We therefore use an intermediate sampling budget in the main configuration, while the larger settings characterize the observed scaling range.

The sample-count sweep compares complete configurations and does not attribute the observed gains solely to training-time sampling. Moreover, the fitted trends describe only the tested configurations; they do not establish a general scaling law or guarantee corresponding online gains. Increasing RM capacity does not increase serving cost because the RM remains offline.

\noindent\textbf{System Optimization.} Table~\ref{tab:systems-efficiency} reports the cumulative effects of five system optimizations under a fixed hardware and software configuration. Training measurements cover decoder step time and peak memory, while serving measurements cover decoder model-path latency and peak memory. This model-path benchmark differs from the end-to-end latency across the reranking and mix-ranking stages reported in Table~\ref{tab:online-ab}.

Rollout compilation removes graph breaks and fuses the autoregressive loop, improving both training and serving efficiency. BF16 further reduces arithmetic and activation costs. KV caching reuses autoregressive states and shares the prefill cache across sampled slates, producing the largest increase in serving speed. Memory-efficient attention further reduces the cost of the long context used during prefill and generation, following FlexAttention~\cite{dong2024flexattention}.

Activation checkpointing applies only to training and trades additional computation for lower memory use. The final training configuration achieves a \(4.91\times\) speedup and an \(81.8\%\) memory reduction relative to eager execution. The final serving configuration achieves a \(16.68\times\) model-path speedup and a \(72.7\%\) memory reduction. These serving measurements cover only decoder execution; the production benchmark covers all processing within the reranking and mix-ranking stages.

Within the tested configurations, additional capacity and sampling compute improve either predictive discrimination or RM-estimated RG. The system results show that cumulative implementation optimizations substantially reduce the corresponding decoder training and serving costs.

\begin{table*}[t]
    \centering
      \caption{Cumulative effects of five system optimizations on OneTrans-GSR decoder training and serving. The listed configurations are cumulative, while all relative
      changes and speedups are computed against the eager baseline. The baseline reports absolute training step time, serving latency, and memory;
      subsequent rows report relative changes and cumulative speedups. Serving latency measures only the decoder model path, whereas Table~\ref{tab:online-ab} reports end-to-end latency across the reranking and mix-ranking stages.}
    \label{tab:systems-efficiency}
    \small
    \setlength{\tabcolsep}{5pt}
    \begin{tabular*}{\textwidth}{@{\extracolsep{\fill}}lrrrrrr@{}}
      \toprule
      & \multicolumn{3}{c}{Training}
      & \multicolumn{3}{c}{Serving} \\
      \cmidrule(lr){2-4}
      \cmidrule(lr){5-7}
      \multirow[c]{2}{*}{Variant}
      & Step Time
      & \multirow[c]{2}{*}{Speedup}
      & Memory
      & Model-Path
      & \multirow[c]{2}{*}{Speedup}
      & Memory \\

      & (ms)
      &
      & (GB)
      & Latency (ms)
      &
      & (GB) \\

      \midrule

      Eager OneTrans-GSR
      & 2452.91
      & \(1.00\times\)
      & 72.7
      & 898
      & \(1.00\times\)
      & 8.1 \\

      + Rollout Compilation
      & \(-43.2\%\)
      & \(1.76\times\)
      & \(-22.4\%\)
      & \(-51.5\%\)
      & \(2.06\times\)
      & \(-7.8\%\) \\

      + BF16
      & \(-56.9\%\)
      & \(2.32\times\)
      & \(-50.8\%\)
      & \(-60.9\%\)
      & \(2.56\times\)
      & \(-25.1\%\) \\

      + KV Cache
      & \(-79.3\%\)
      & \(4.85\times\)
      & \(-50.7\%\)
      & \(-91.3\%\)
      & \(11.57\times\)
      & \(-35.5\%\) \\

      + Memory-Efficient Attention
      & \(\mathbf{-80.6\%}\)
      & \(\mathbf{5.19\times}\)
      & \(-76.4\%\)
      & \(\mathbf{-93.9\%}\)
      & \(\mathbf{16.68\times}\)
      & \(\mathbf{-72.7\%}\) \\

      + Activation Checkpointing
      & \(-79.5\%\)
      & \(4.91\times\)
      & \(\mathbf{-81.8\%}\)
      & --
      & --
      & -- \\

      \bottomrule
    \end{tabular*}
  \end{table*}

\subsection{RQ6: Production A/B Validation}
\label{sec:online}

We assess the business impact of OneTrans-GSR with a randomized online A/B test in the production system described in Section~\ref{sec:setup}. The control group is served by the deployed Generator--Evaluator system, including its constrained generator, slate-level evaluator, and business rules. In the treatment group, OneTrans-GSR replaces the reranking and mix-ranking cascade. The upstream retrieval and fine-ranking stages and the business rule set are held fixed across the two groups.

Users are randomly assigned at the account level, with equal shares of eligible traffic allocated to each group during the week-long experiment. The deployed policy combines Separate Standardization during post-training with constrained decoding and serves the generated slate directly. The experiment evaluates the complete replacement of the G-E cascade; it does not isolate the contribution of individual components.

The evaluation considers GMV/u, advertiser value, repurchase rate, and serving latency. Serving latency is measured end-to-end across the reranking and mix-ranking stages, excluding upstream retrieval and fine ranking. The repurchase rate is evaluated only after the complete 14-day observation window has elapsed.

Relative to the G-E control, OneTrans-GSR increases GMV/u by \(2.63\%\),
advertiser value by \(2.06\%\), and repurchase rate by \(2.03\%\), with all
three reported gains significant at \(p<0.05\)
(Table~\ref{tab:online-ab}). End-to-end latency across the reranking and
mix-ranking stages decreases by \(13\%\). These results provide independent production evidence for the
complete system in this setting: offline slate scoring can guide a single
autoregressive serving policy that improves measured business outcomes and
reduces serving latency.

\begin{table}[t]
\centering
\caption{Production A/B results for OneTrans-GSR relative to the G-E control. Serving latency is measured end-to-end across the reranking and mix-ranking stages. Arrows indicate the preferred direction, and \(^{*}\) denotes \(p<0.05\).}
\label{tab:online-ab}
\small
\setlength{\tabcolsep}{8pt}
\begin{tabular}{@{}lc@{}}
  \toprule
  Metric & Relative Change \\
  \midrule
  GMV/u $\uparrow$                       & $+2.63\%^{*}$ \\
  Advertiser Value $\uparrow$            & $+2.06\%^{*}$ \\
  Repurchase Rate $\uparrow$             & $+2.03\%^{*}$ \\
  Serving Latency $\downarrow$           & $-13\%$ \\
  \bottomrule
\end{tabular}
\end{table}

\FloatBarrier
\section{Conclusion}
\label{sec:conclusion}

We present OneTrans-GSR, a single autoregressive policy for reranking and mix-ranking over a request-specific candidate pool. The decoder selects candidates, allocates exposure across content types, and determines their order through one generation process. A slate-level RM learns from logged user feedback and provides rewards for click, order, and GMV objectives, complemented by a delayed repurchase proxy for LTV. GMR-PO aggregates these objectives into a scalar reward and combines the standardized RM advantage with a separately standardized advantage derived from soft business preferences. Hard masks enforce strict requirements, while finite logit adjustments provide soft control during decoding. Because the RM is used only for offline post-training, its capacity can scale independently of the online serving model.

In the evaluated production system, OneTrans-GSR replaces the Generator--Evaluator cascade. Offline experiments show positive RM-estimated utility gains relative to logged G-E slates. The generated slates also exhibit greater diversity and lower redundancy and fatigue than score-based ranking baselines, including when additive logit adjustments are disabled. The scaling experiments show improved RM discrimination or RM-estimated policy RG across the tested capacity and sampling configurations. In the randomized production A/B test, OneTrans-GSR increases GMV/u by \(2.63\%\), advertiser value by \(2.06\%\), and repurchase rate by \(2.03\%\), while reducing end-to-end latency across the reranking and mix-ranking stages by \(13\%\).

Although evaluated in a single production system, OneTrans-GSR is based on a general formulation: generating an ordered slate from a request-specific candidate pool using slate-level rewards and constrained autoregressive decoding. Different applications may define their own inputs, objectives, and constraints, while the current production experiment validates the resulting system as a whole rather than attributing its gains to individual components.

This formulation applies broadly to recommendation, search, advertising, and other settings that construct ordered slates under multiple objectives and constraints. By keeping slate evaluation offline and serving a single policy that generates the final slate directly, OneTrans-GSR provides a scalable framework for learning these decisions from data while retaining explicit business control. Its ability to benefit from additional reward-model, policy, and sampling compute suggests a promising direction for learned slate generation beyond e-commerce reranking and mix-ranking.

\clearpage
\begingroup
\raggedright
\bibliographystyle{plainnat}
\bibliography{references}
\endgroup
\appendix
\section{Full Author List}
\noindent
\textbf{Foundation Team:}
Feng Cheng\textsuperscript{*},
Yu-Hsuan Huang\textsuperscript{*},
Fengyi Wang\textsuperscript{*},
Chaofan Zhou,
Zhennan Zhu,
Huan He,
Bo Zhang,
Ruoqi Zhang,
Zhanfeng Mo,
Ziwei Liu,
Lanyi Li,
Shaowei Liu\textsuperscript{\textdagger}.

\noindent
\textbf{Architecture Team:}
Yifeng Pan,
Junhua Long,
Ziqiu Zhou,
Yue Lin.

\smallskip
\noindent
\textsuperscript{*}Equal contribution. \textsuperscript{\textdagger}Corresponding author: \href{mailto:liushaowei.nphard@bytedance.com}{liushaowei.nphard@bytedance.com}.
\end{document}